\documentclass[a4paper,11pt]{article}
\usepackage[utf8]{inputenc}
\usepackage[T1]{fontenc}
\usepackage[english]{babel}
\usepackage[hmargin={32mm,32mm},vmargin={32mm,35mm}]{geometry}
\usepackage{amsmath}
\usepackage{amsfonts}
\usepackage{amssymb}
\usepackage{enumerate}
\usepackage{graphicx}
\usepackage{tocloft}
\usepackage{bbm}
\usepackage[dvipsnames]{xcolor}
\usepackage{xspace}
\usepackage{url}
\usepackage[hidelinks]{hyperref}
\usepackage{cite}
\usepackage{listings}
\usepackage{caption}
\usepackage{subcaption}
\usepackage{empheq}

\usepackage{amstext} 
\usepackage{array}   
\newcolumntype{C}{>{$}c<{$}} 

\usepackage{bookmark}

\newcommand{\ket}[1]{|#1\rangle}
\newcommand{\braket}[2]{\langle #1 | #2 \rangle}

\newcommand{\be}{\begin{equation}}
\newcommand{\ee}{\end{equation}}
\newcommand{\bea}{\begin{eqnarray}}
\newcommand{\eea}{\end{eqnarray}}
\newcommand{\bal}{\begin{align}}
\newcommand{\eal}{\end{align}}

\newcommand{\eg}{e.g.\@\xspace}
\newcommand{\ie}{i.e.\@\xspace}
\newcommand{\Eq}[1]{Eq.\@\xspace\eqref{#1}}
\newcommand{\Eqs}[1]{Eqs.\@\xspace\eqref{#1}}
\newcommand{\Fig}[1]{Fig.\@\xspace\ref{#1}}

\newcommand{\updown}[2]{^{#1}_{\phantom{#1}#2}}
\newcommand{\downup}[2]{_{#1}^{\phantom{#1}#2}}

\DeclareMathOperator{\Tr}{{\rm Tr}}

\newcommand{\Id}{\mathbbm{1}}

\numberwithin{equation}{section}

\newsavebox{\mybox}

\newcommand{\Rt}{\,{}^{(3)}\!R}
\newcommand{\Et}{\widetilde E}
\newcommand{\EEt}{\widetilde{\cal E}}
\newcommand{\de}{|{\rm det}\,E|}
\newcommand{\dE}[1]{\bigl|{\rm det}\,E(#1)\bigr|}
\newcommand{\Gc}{{\Gamma_{\! 0}}}

\newcommand{\A}{{\mathbf A}}
\newcommand{\B}{{\mathbf B}}
\newcommand{\D}{{\cal D}}

\begin{document}

\begin{center}

\Large
\textbf{Scalar curvature operator for models of \\
loop quantum gravity on a cubical graph}

\vspace{16pt}

\large
Jerzy Lewandowski and Ilkka Mäkinen

\vspace{8pt}

\normalsize
Faculty of Physics, University of Warsaw \\ 
Pasteura 5, 02-093 Warsaw, Poland \\
jerzy.lewandowski@fuw.edu.pl, ilkka.makinen@fuw.edu.pl 

\end{center}

\renewcommand{\abstractname}{\vspace{-\baselineskip}}

\begin{abstract}

\noindent In this article we introduce a new operator representing the three-dimensional scalar curvature in loop quantum gravity. Our construction does not apply to the entire kinematical Hilbert space of loop quantum gravity; instead, the operator is defined on the Hilbert space of a fixed cubical graph. The starting point of our work is to write the spatial Ricci scalar classically as a function of the densitized triad and its $SU(2)$-covariant derivatives. We pass from the classical expression to a quantum operator through a regularization procedure, in which covariant derivatives of the triad are discretized as finite differences of gauge covariant flux variables on the cubical lattice provided by the graph. While more work is needed in order to extend our construction to encompass states based on all possible graphs, the operator presented here can be applied in models such as quantum-reduced loop gravity and effective dynamics, which are derived from the kinematical framework of full loop quantum gravity, and are formulated in terms of states defined on cubical graphs.

\end{abstract}

\section{Introduction}

Loop quantum gravity \cite{lqg0, lqg1, lqg2, lqg3} is one of the main candidates for a quantum theory of gravitation. The quantum states of the gravitational field in loop quantum gravity -- the well-known spin network states -- have a natural physical interpretation as states describing discrete, quantized spatial geometries \cite{discreteness}. Thus, loop quantum gravity provides a concrete realization of a quantum theory of gravity as a theory of quantum geometry.

Geometrical observables are represented in loop quantum gravity as self-adjoint operators acting on the kinematical Hilbert space of the theory. Operators corresponding to areas of surfaces and volumes of spatial regions \cite{discreteness, area, volume} are of essential importance in loop quantum gravity, the fundamentally discrete nature of geometry being encoded in the discrete spectra of these basic geometric operators. Another key example of a geometrical observable is the Ricci scalar of the spatial manifold (in the context of a 3+1 formulation of general relativity). In addition to being an important geometrical quantity characterizing the geometry of the spatial manifold, the Ricci scalar appears in the Hamiltonian constraint of general relativity, and hence has a direct connection with the dynamics of both classical general relativity and loop quantum gravity.

An operator representing the scalar curvature of the spatial manifold has been constructed for loop quantum gravity in \cite{curvature}. The construction is based on the observation that the integral of the Ricci scalar over the spatial manifold,
\begin{equation}
	\int d^3x\,\sqrt q\Rt,
	\label{integral}
\end{equation}
is also the Einstein--Hilbert action for Euclidean general relativity in three dimensions. Thus, using the ideas of Regge calculus \cite{Regge}, the integral can be expressed in terms of the hinge lengths and deficit angles associated to a cellular decomposition of the spatial manifold. This fact, together with the knowledge that operators representing lengths and angles \cite{Bianchi, Major} are readily available in loop quantum gravity, enables one to promote the integral \eqref{integral} into a well-defined operator of loop quantum gravity in a rather straightforward fashion.

Nevertheless, due to the indirect nature of the strategy of turning to Regge calculus in order to quantize the integrated Ricci scalar, it is not obvious whether the properties of the resulting operator are fully satisfactory from the physical point of view. For instance, the curvature operator can be used as a part of the Hamiltonian constraint operator, to replace the Lorentzian part of the constraint commonly used in loop quantum gravity. Then it seems unclear whether the operator of \cite{curvature} can be consistent with any operator proposed so far as a quantization of the Euclidean part of the constraint. A conflict could conceivably arise due to the fact that the classical foundations underlying all available quantizations of the Euclidean part refer to the smooth, continuous physical geometry of the spatial manifold, whereas Regge's formula for the integral \eqref{integral} instead makes use of an auxiliary manifold whose geometry is singular, with curvature being concentrated entirely on one-dimensional submanifolds.

In this article we propose a new operator representing the integrated scalar curvature \eqref{integral}. The starting point of our construction is to write the Ricci scalar of the spatial manifold in a suitable way as a function of the Ashtekar variables, after which a curvature operator can be defined through a regularization procedure whereby the integral \eqref{integral} is expressed in terms of quantities corresponding to well-defined operators in loop quantum gravity. On the technical level, the main challenge encountered in such an approach is to find an appropriate regularization of spatial derivatives of the densitized triad. Our method of dealing with this challenge consists of two key steps. Firstly, we take as our classical starting point an expression which gives the Ricci scalar as a function of the densitized triad and its $SU(2)$-covariant derivatives, as opposed to partial derivatives. Secondly, we will not attempt to define a curvature operator in the entire Hilbert space of loop quantum gravity, instead limiting our considerations to states defined on a fixed cubical graph\footnote{By a cubical graph we mean a graph whose nodes are six-valent, and whose edges are aligned with the coordinate directions defined by a fiducial Cartesian background coordinate system.}. Under this restriction, it becomes relatively straightforward to regularize gauge covariant derivatives by discretizing them as finite differences of so-called parallel transported flux variables on the rectangular lattice provided by the graph.

The restriction to a fixed cubical graph implies that our construction does not apply to a vast majority of the states which span the kinematical Hilbert space of loop quantum gravity. In the context of full loop quantum gravity, a natural framework for interpreting the operator proposed in this article is provided by algebraic quantum gravity \cite{aqg1, aqg4} -- an approach which uses the mathematical formalism of loop quantum gravity to accomplish a quantization of the full set of gravitational degrees of freedom, and which is formulated entirely in terms of states defined on a single (abstract, algebraic) cubical graph. From the point of view of loop quantum gravity in its standard formulation, where the Hilbert space of the theory includes states based on all possible graphs, the construction presented in this article is best seen as a preliminary investigation, whose ideas and techniques may -- with more work -- eventually be extended to yield a well-defined curvature operator on the entire Hilbert space of loop quantum gravity.

On the other hand, our work in its present form is directly relevant to several physically motivated models of loop quantum gravity, in which states based on cubical graphs are used to perform practical calculations. Models of this kind include quantum-reduced loop gravity \cite{qrlg1, qrlg4, operators} -- a simplified model of loop quantum gravity, which is derived from the full theory through a procedure representing a gauge fixing of the densitized triad -- and the effective dynamics approach \cite{effective1, effective2, effective3}, where the expectation value of the Hamiltonian operator with respect to a family of semiclassical states is considered as an effective Hamiltonian function generating evolution on a classical phase space. For such models, our construction provides a well-defined curvature operator which is ready to be used in concrete applications, and whose physical properties may be more satisfactory than those of the Regge calculus -inspired curvature operator introduced in \cite{curvature}.\footnote{In the companion article \cite{part2} we examine the operator proposed in the present article in the setting of quantum-reduced loop gravity, following the point of view established in \cite{operators}, where it was shown that operators of the quantum-reduced model are obtained from the corresponding operators of full loop quantum gravity simply by letting the operators of the full theory act on states in the Hilbert space of quantum-reduced loop gravity and discarding terms of subleading order in the spin quantum number $j$. We find that our operator yields a non-trivial and seemingly adequate curvature operator for the quantum-reduced model, unlike the operator of \cite{curvature}, whose action on any state in the quantum-reduced Hilbert space is trivially vanishing.}

The material in this article is organized as follows. The present introductory section is followed by section 2, where we introduce the expression which gives the three-dimensional Ricci scalar as a function of the densitized triad and its gauge covariant derivatives, and which forms the classical starting point of our work. This section also serves to fix our notation and conventions regarding the basic notions of the Riemannian geometry of the spatial manifold. In section 3 we give a brief review of the key elements of loop quantum gravity, focusing particularly on those aspects which will play a role in our construction of the curvature operator. The construction itself is then presented in the following two sections. Apart from the restriction concerning the graph, the construction is completely general, and is not tied to any particular model of loop quantum gravity which makes use of states based on cubical graphs. In section 4 we perform the regularization of the classical Ricci scalar on a cubical graph, and in section 5 we carry out the quantization of the regularized expression and discuss the main features of the resulting curvature operator. Finally, in section 6 we conclude by summarizing and assessing our results. The article also includes two appendices, in which we provide a derivation of the classical identity expressing the Ricci scalar in terms of the Ashtekar variables, and verify the validity of the regularized expressions introduced in section 4 to approximate covariant derivatives of the triad.

\section{The classical Ricci scalar}

The classical setting for our work is the 3+1 formulation of general relativity \cite{ADM1, ADM2}, expressed in terms of the Ashtekar variables \cite{Ashtekar1986, Ashtekar1987, Barbero}. The elementary variables of the formalism are the Ashtekar--Barbero connection $A_a^i$ and its conjugate, the densitized triad $E^a_i$. The densitized triad is related to the inverse metric of the spatial manifold $\Sigma$ according to
\begin{equation}
	q^{ab} = \frac{E^a_iE^b_i}{\de},
	\label{q^ab}
\end{equation}
where $\det E \equiv \det E^a_i$ is the determinant of the densitized triad. The metric itself can be written as
\begin{equation}
	q_{ab} = \de E_a^iE_b^i,
	\label{q_ab}
\end{equation}
where
\begin{equation}
	E_a^i \equiv \frac{1}{2\det E}\epsilon_{abc}\epsilon^{ijk}E^b_jE^c_k.
	\label{}
\end{equation}
is the inverse of the densitized triad, both with respect to the internal index and the spatial index:
\begin{equation}
	E^a_iE^i_b = \delta^a_b, \qquad E^a_iE^j_a = \delta_i^j.
	\label{}
\end{equation}
Note that $E^a_i$ is a density of weight $1$, whereas $E_a^i$ is a density of weight $-1$. Accordingly, $E_a^i$ is not obtained from $E^a_i$ simply by using the metric $q_{ab}$ to lower the spatial index; rather, the relation between the densitized triad and its inverse reads $E_a^i = q_{ab}E^b_i/\de$.

We define the Riemann tensor on $\Sigma$ according to the convention
\begin{equation}
	[D_a,D_b]v^c = \Rt\updown{c}{dab}v^d,
	\label{}
\end{equation}
where $D_a$ is the covariant derivative compatible with the metric $q_{ab}$. Under this definition, the Riemann tensor is given by the expression
 \begin{equation}
	 \Rt\updown{a}{bcd} = \partial_c\Gamma^a_{bd} - \partial_d\Gamma^a_{bc} + \Gamma^a_{ce}\Gamma^e_{bd} - \Gamma^a_{de}\Gamma^e_{bc},
	\label{}
\end{equation}
where $\Gamma^a_{bc}$ are the Christoffel symbols corresponding to the spatial metric. We then have the Ricci tensor
\begin{equation}
	\Rt_{ab} = \Rt\updown{c}{acb}
	\label{}
\end{equation}
and the Ricci scalar
\begin{equation}
	\Rt = q^{ab}\Rt_{ab} =  q^{ab}\bigl(\partial_c\Gamma^c_{ab} - \partial_b\Gamma^c_{ac} + \Gamma^c_{ab}\Gamma^d_{cd} - \Gamma^c_{ad}\Gamma^d_{bc}\bigr).
	\label{R(q)}
\end{equation}
\Eq{R(q)} expresses the Ricci scalar as a function of the spatial metric $q_{ab}$. To derive an expression for the Ricci scalar in terms of the densitized triad, it suffices to insert \Eqs{q^ab} and \eqref{q_ab} into \Eq{R(q)} and evaluate all the resulting derivatives. The calculation, which is somewhat lengthy but in principle straightforward, is outlined in Appendix \ref{sec:A} and yields the result

\begin{align}
	\de\Rt = & -2E^a_i\partial_a\partial_bE^b_i + 2Q^{ab}E_c^i\partial_a\partial_bE^c_i \notag \\
	& - (\partial_a E^a_i)(\partial_bE^b_i) - \frac{1}{2}(\partial_aE^b_i)(\partial_bE^a_i) \notag \\
	& + \frac{5}{2}Q^{ab}(\partial_aE^c_i)(\partial_bE_c^i) - \frac{1}{2}Q^{ab}Q_{cd}(\partial_aE^c_i)(\partial_bE^d_i) \notag \\
	& + 2A\updown{ab}{a}B\downup{cb}{c} + 2A\updown{ab}{b}B\downup{ca}{c} + A\updown{ab}{c}B\downup{ba}{c} \notag \\
	& + \frac{1}{2}Q_{ab}A\updown{ca}{d}A\updown{db}{c} - Q^{ab}B\downup{ca}{c}B\downup{db}{d} \notag \\
	& + 2\bigl(Q^{ab}B\downup{ca}{c} - A\updown{ab}{a} - A\updown{ba}{a}\bigr)C_b + \frac{3}{2}Q^{ab}C_aC_b - 2Q^{ab}C_{ab}
	\label{qR_v1}
\end{align}
where we have introduced the following abbreviations:
\begin{align}
	Q^{ab} &= E^a_i E^b_i \label{Q^ab} \\
	Q_{ab} &= E_a^i E_b^i \label{Q_ab} \\
	A\updown{ab}{c} &= E^a_i\partial_c E^b_i \\
	B\downup{ab}{c} &= E_a^i\partial_b E^c_i \\
	C_a &= \frac{\partial_a\de}{\de} \label{C_a} \\
	C_{ab} &= \frac{\partial_a\partial_b\de}{\de}. \label{C_ab}
\end{align}

From the perspective of promoting the expression \eqref{qR_v1} into an operator representing the Ricci scalar in loop quantum gravity, the problematic feature are the partial derivatives of the densitized triad. While in principle it is possible to construct operators corresponding to partial derivatives of the triad, it seems unclear how this approach could result in a gauge invariant\footnote{In the sense of internal $SU(2)$ gauge transformations associated with local rotations of the triad.} curvature operator. An alternative expression for the Ricci scalar as a function of the Ashtekar variables, which provides a more suitable classical starting point for the construction of the curvature operator, can be obtained by using, instead of partial derivatives of the densitized triad, its gauge covariant derivatives defined by
\begin{equation}
	\D_aE^b_i = \partial_aE^b_i + \epsilon\downup{ij}{k}A_a^jE^b_k.
	\label{}
\end{equation}
Equivalently, using the $su(2)$-valued variables $E^a = E^a_i\tau^i$ and $A_a = A_a^i\tau_i$, we can write the definition as
\begin{equation}
	\D_aE^b = \partial_aE^b + \bigl[A_a, E^b\bigr].
	\label{DE}
\end{equation}
Under a local $SU(2)$ gauge transformation given by a gauge function $g(x) \in SU(2)$, the gauge covariant derivative transforms covariantly, \ie
\begin{equation}
	\D_aE^b(x) \to g(x)\D_aE^b(x)g^{-1}(x).
	\label{g*DE}
\end{equation}
Due to this property, we can apply the definition \eqref{DE} to the covariant derivative itself, obtaining the expression
\begin{align}
	\D_a\D_bE^c = \partial_a\partial_bE^c + \bigl[A_b, \partial_aE^c\bigr] + \bigl[A_a, \partial_bE^c\bigr] + \bigl[\partial_aA_b, E^c\bigr] + \Bigl[A_a, \bigl[A_b, E^c\bigr]\Bigr]
	\label{DDE}
\end{align}
for the second covariant derivative of the triad.

Now we can use \Eqs{DE} and \eqref{DDE} to replace all partial derivatives of the triad in \Eq{qR_v1} with covariant derivatives. The calculation, which is briefly discussed in section \ref{sec:replacing}, shows that the correction terms generated by this replacement cancel out among themselves, provided that one uses the symmetric part of the second covariant derivative $\D_a\D_bE^c_i$ to replace the second partial derivative $\partial_a\partial_bE^c_i$, which is symmetric in $a$ and $b$. Note also that no correction terms arise from the factors involving derivatives of $\de$, which is gauge invariant, and therefore its gauge covariant derivative is simply identical with the partial derivative.

Thus, our conclusion is that the classical Ricci scalar can be expressed in the alternative form
\begin{align}
	\de\Rt = & -2E^a_i\D_{(a}\D_{b)}E^b_i + 2Q^{ab}E_c^i\D_a\D_bE^c_i \notag \\
	& - (\D_a E^a_i)(\D_bE^b_i) - \frac{1}{2}(\D_aE^b_i)(\D_bE^a_i) \notag \\
	& + \frac{5}{2}Q^{ab}(\D_aE^c_i)(\D_bE_c^i) - \frac{1}{2}Q^{ab}Q_{cd}(\D_aE^c_i)(\D_bE^d_i) \notag \\
	& + 2\A\updown{ab}{a}\B\downup{cb}{c} + 2\A\updown{ab}{b}\B\downup{ca}{c} + \A\updown{ab}{c}\B\downup{ba}{c} \notag \\
	& + \frac{1}{2}Q_{ab}\A\updown{ca}{d}\A\updown{db}{c} - Q^{ab}\B\downup{ca}{c}\B\downup{db}{d} \notag \\
	& + 2\bigl(Q^{ab}\B\downup{ca}{c} - \A\updown{ab}{a} - \A\updown{ba}{a}\bigr)C_b + \frac{3}{2}Q^{ab}C_aC_b - 2Q^{ab}C_{ab}, \label{qR}
\end{align}
where the new abbreviations
\begin{align}
	\A\updown{ab}{c} &= E^a_i\D_c E^b_i \label{Aabc} \\
	\B\downup{ab}{c} &= E_a^i\D_b E^c_i \label{Babc} 
\end{align}
have been introduced. Since the covariant derivatives of the triad transform under gauge transformations according to \Eq{g*DE}, the expression \eqref{qR} for the Ricci scalar is manifestly gauge invariant, in contrast to \Eq{qR_v1}, whose gauge invariance is not immediately apparent.

\Eq{qR} forms the classical starting point for our construction of an operator representing the integrated Ricci scalar
\begin{equation}
	\int d^3x\,\sqrt q\Rt.
	\label{int R}
  \end{equation}
Besides being an essential geometrical observable in its own right, the Ricci scalar also plays a role in the formulation of the dynamics of loop quantum gravity (see section \ref{sec:dynamics}). For the latter purpose, one needs an operator representing the Ricci scalar integrated against an arbitrary smearing function $N(x)$,
\begin{equation}
	\int d^3x\,N\sqrt q\Rt.
	\label{int NR}
\end{equation}
The construction presented in this article also provides a well-defined operator corresponding to the smeared Ricci scalar \eqref{int NR}.

\section{Loop quantum gravity}

In this section we will briefly review the basic elements of the kinematical structure of loop quantum gravity, focusing on those aspects of the framework which are relevant to the work presented in this article. We will recall the kinematical Hilbert space of loop quantum gravity and the basic operators of the theory. In particular, we will introduce and establish the basic properties of the so-called parallel transported flux operator, which is a key ingredient in our construction of the curvature operator. For a more detailed presentation of the foundations of loop quantum gravity, we refer the reader \eg to \cite{lqg0, lqg1, lqg2, lqg3, lqg4, lqg5, thesis}.

\subsection{The kinematical Hilbert space}

The kinematical Hilbert space of loop quantum gravity is formed by the so-called cylindrical functions. A cylindrical function is labeled by a graph $\Gamma$, with edges $e_1, \dots, e_N$ (which are assumed to be oriented, and embedded in the spatial manifold $\Sigma$). A function cylindrical with respect to a graph $\Gamma$ is essentially a complex-valued function of the form
\begin{equation}
	\Psi_\Gamma(h_{e_1},\dots,h_{e_N}),
	\label{}
\end{equation}
where the arguments of the function are $SU(2)$ group elements, one for each edge of the graph. The group elements $h_e$ are usually referred to as holonomies, due to their classical origin as holonomies of the Ashtekar--Barbero connection.

The holonomies are assumed to satisfy certain algebraic properties, reflecting their role as parallel transport operators in the classical theory. We have
\begin{equation}
	h_{e^{-1}} = h_e^{-1}
\end{equation}
where $e^{-1}$ denotes $e$ taken with the opposite orientation;
\begin{equation}
	h_{e_2}h_{e_1} = h_{e_2\circ e_1}
	\label{}
\end{equation}
where the endpoint of $e_1$ coincides with the beginning point of $e_2$, and $e_2\circ e_1$ denotes the edge obtained by joining $e_1$ and $e_2$;
and
\begin{equation}
	h_p = \Id
	\label{}
\end{equation}
if $p$ is a curve consisting of a single point.

A scalar product on the space of cylindrical functions can be defined in a natural way using the Haar measure of $SU(2)$. For two cylindrical functions based on the same graph $\Gamma$, one defines
\begin{equation}
	\braket{\Psi_\Gamma}{\Phi_\Gamma} = \int dg_1\cdots dg_N\,\overline{\Psi_\Gamma(g_1,\dots,g_N)}\Phi_\Gamma(g_1,\dots,g_N),
	\label{scalar}
\end{equation}
where $dg$ is the normalized Haar measure of $SU(2)$. The definition is extended to cylindrical functions based on two different graphs $\Gamma_1$ and $\Gamma_2$ by taking any larger graph $\Gamma_{12}$ that contains $\Gamma_1$ and $\Gamma_2$ as subgraphs, and viewing the functions as cylindrical functions on $\Gamma_{12}$ in the standard way (by introducing a trivial dependence on the group elements associated with the additional edges, see \eg \cite{lqg1}) and then applying \Eq{scalar}.  

A basis of the space of cylindrical functions can be constructed using the $SU(2)$ representation matrices $D^{(j)}_{mn}(g)$. By the Peter--Weyl theorem, the space of functions cylindrical with respect to a graph $\Gamma$ is spanned by the functions
\begin{equation}
	\prod_{e\in\Gamma} D^{(j_e)}_{m_en_e}(h_e),
	\label{basis_kin}
\end{equation}
as the quantum numbers $j_e$, $m_e$ and $n_e$ range over their possible values. The basis states \eqref{basis_kin} are orthogonal but not normalized under the scalar product \eqref{scalar}. In order to obtain a normalized basis, each representation matrix in \Eq{basis_kin} should be multiplied with the factor $\sqrt{d_{j_e}}$, where $d_j\equiv 2j+1$ denotes the dimension of the spin-$j$ representation of $SU(2)$.

At the classical level, the Ashtekar formulation of general relativity enjoys a local $SU(2)$ gauge symmetry, which corresponds to rotations of the densitized triad with respect to the internal index (the spatial metric being invariant under such rotations). These gauge transformations are generated in the classical theory by the Gauss constraint
\begin{equation}
	G_i = \partial_aE^a_i + \epsilon\downup{ij}{k}A_a^jE^a_k.
	\label{}
\end{equation}
Within the kinematical Hilbert space of loop quantum gravity, one can identify the subspace consisting of states which are invariant under the analogous gauge transformations in the quantum theory.

Under a local gauge transformation described by a gauge function $g(x)\in SU(2)$, the holonomy associated to an edge $e$ transforms as
\begin{equation}
	h_e \to g(t_e)h_eg^{-1}(s_e),
	\label{gauge}
\end{equation}
where $s_e$ and $t_e$ denote the beginning point and endpoint (''source'' and ''target'') of $e$. From this one deduces that the space of functions cylindrical with respect to a graph $\Gamma$, and invariant under $SU(2)$ gauge transformations, is spanned by functions of the form
\begin{equation}
	\biggl(\prod_{v\in\Gamma} \iota_v\biggr)\cdot\biggl(\prod_{e\in\Gamma} D^{(j_e)}(h_e)\biggr).
	\label{basis_g}
\end{equation}
Here an invariant tensor $\iota_v$ -- usually referred to as an intertwiner -- is assigned to each node $v$ of the graph, and the dot symbolizes a complete contraction of magnetic indices in the way indicated by the pattern of the graph. The condition for $\iota_v$ to be invariant has to be understood in the appropriate sense, taking into account the orientation of the graph. If the node $v$ contains $M$ incoming edges (carrying spins $j_1, \dots, j_M$) and $N-M$ outgoing edges (carrying spins $j_{M+1}, \dots, j_N$), this condition reads 
\begin{equation}
	D^{(j_1)}(g)\cdots D^{(j_M)}(g)D^{(j_{M+1})}(g^{-1})\cdots D^{(j_N)}(g^{-1})\iota_v = \iota_v
	\label{}
\end{equation}
with $\iota_v$ being viewed as a tensor having $M$ upper indices and $N-M$ lower indices, and the representation matrices acting on $\iota_v$ by contraction of magnetic indices.

\subsection{Kinematical operators}
The elementary operators of loop quantum gravity are the holonomy and flux operators. The holonomy operator $D^{(j)}_{mn}(h_e)$ acts on cylindrical functions by multiplication. The form of the resulting state,
\begin{equation}
	D^{(j)}_{mn}(h_e)\Psi_\Gamma(h_{e_1},\dots,h_{e_N}),
	\label{Dpsi}
\end{equation}
depends on whether the edge $e$ is contained among the edges $e_1, \dots, e_N$. If $e$ is not an edge of $\Gamma$, the state \eqref{Dpsi} defines a cylindrical function on the graph $\Gamma\cup e$. If $e$ coincides with an edge of $\Gamma$, the function \eqref{Dpsi} is still a cylindrical function on the graph $\Gamma$. In the latter case, if the state $\Psi_\Gamma(h_{e_1},\dots,h_{e_N})$ is given in the basis \eqref{basis_kin} or \eqref{basis_g}, the result of the multiplication \eqref{Dpsi} can be expressed in the same basis by coupling the holonomies on the edge $e$ by means of the Clebsch--Gordan series
\begin{equation}
	D^{(j_1)}_{m_1n_1}(h_e)D^{(j_2)}_{m_2n_2}(h_e) = \sum_j C^{(j_1\;j_2\;j)}_{m_1\;m_2\;m_1+m_2} C^{(j_1\;j_2\;j)}_{n_1\;n_2\;n_1+n_2}D^{(j)}_{m_1+m_2\;n_1+n_2}(h_e),
	\label{DD}
\end{equation} 
where $C^{(j_1\;j_2\;j)}_{m_1\;m_2\;m}$ are the Clebsch--Gordan coefficients of $SU(2)$. 

The flux operator is a quantization of the classical variable
\begin{equation}
	E_i(S) = \int d^2\sigma\,n_a(\sigma)E^a_i\bigl(x(\sigma)\bigr)
	\label{}
\end{equation}
representing the flux of the densitized triad through the surface $S$. If we consider just one edge, which has a single intersection with the surface $S$ at a point $v$, the action of the flux operator reads
\begin{equation}
	E_i(S)D^{(j)}(h_e) = i\nu(S, e)\times\begin{cases}
	\dfrac{1}{2}D^{(j)}(h_e)\tau_i^{(j)} & \text{if $e$ begins from $v$} \\[1.7ex] 
		\dfrac{1}{2}\tau_i^{(j)} D^{(j)}(h_e) & \text{if $e$ ends at $v$} \\[1.7ex]
		D^{(j)}(h_{e_1})\tau_i^{(j)}D^{(j)}(h_{e_2}) & \text{if $v$ is an interior point of $e$}
	\end{cases}
	\label{Epsi}
\end{equation}
Here $\tau_i^{(j)}$ are the anti-Hermitian generators of $SU(2)$ in the spin-$j$ representation, and $\nu(S, e)$ denotes the relative orientation of $S$ and $e$, \ie $\nu(S,e) = +1$ or $\nu(S,e) = -1$ according to whether the orientation of $e$ at the intersection point agrees with or is opposite to the orientation of $S$, and $\nu(S,e) = 0$ if the edge intersects the surface tangentially. When the flux operator is applied on a cylindrical function, its action obeys the Leibniz rule, in the sense that each intersection between the surface and an edge contributes a term of the form \eqref{Epsi}.

Operators representing other classical quantities can be constructed by expressing the classical function in terms of the elementary variables, \ie holonomies and fluxes, and then promoting the resulting expression into an operator. A basic example of an operator of this kind is the volume operator \cite{volume}, which is the quantization of the classical observable
\begin{equation}
	\int d^3x\,\sqrt{\de}.
	\label{}
\end{equation}
The action of the volume operator on a state based on a graph $\Gamma$ takes the form\footnote{
	Strictly speaking, the definition of the operator $q_v$ involves an undetermined multiplicative factor $\kappa_0$, which arises when one performs an averaging over the background structures used in the construction in order to ensure that the volume operator transforms covariantly under diffeomorphisms \cite{volume}. In this work we use the value $\kappa_0 = 1/48$ for this factor. This choice is justified by a calculation presented in the companion paper \cite{part2}. Essentially, it is the unique value of $\kappa_0$ for which the operator representing the regularized inverse triad \eqref{E_a^i reg} behaves as the inverse of the flux operator \eqref{Epsi} in the Hilbert space of quantum-reduced loop gravity. This agrees with the value originally found by Thiemann and Giesel in \cite{consistency1, consistency2} through a different kind of consistency argument.
}
\begin{equation}
	V\ket{\Psi_\Gamma} = \sum_{v\in\Gamma} \sqrt{|q_v|}\,\ket{\Psi_\Gamma},
	\label{Vpsi}
\end{equation}
where the operator $q_v$ can be expressed in terms of the left- and right-invariant vector fields of $SU(2)$ (an explicit definition of $q_v$ can be found \eg in \cite{volume}), and the sum receives contributions only from nodes of valence three or higher\footnote{In the gauge-invariant subspace formed by the states \eqref{basis_g}, the action of the volume operator on a three-valent node vanishes identically. However, if applied to a generic, non-gauge invariant state, the volume operator generally has a non-zero action also on three-valent nodes.}.

\subsection{Parallel transported flux operator}
\label{sec:transported}

The parallel transported flux operator (also often referred to as the gauge covariant flux in the literature) is a useful modification of the standard flux operator defined by \Eq{Epsi}. The operator is a quantization of the classical function
\begin{equation}
	\Et_i(S, x_0) = -2\,{\rm Tr}\,\Bigl(\tau_i \Et(S, x_0)\Bigr)
	\label{Et_i}
\end{equation}
where $\Et(S, x_0)$ is the matrix-valued variable 
\begin{equation}
	\Et(S, x_0) = \int_S d^2\sigma\,n_a(\sigma)h_{x_0, x(\sigma)}E^a\bigl(x(\sigma)\bigr)h^{-1}_{x_0, x(\sigma)}.
	\label{Et}
\end{equation}
Here $E^a \equiv E^a_i\tau^i$, and $h_{x_0, x(\sigma)} \equiv h_{p_{x_0, x(\sigma)}}$ are holonomies which connect each point $x(\sigma)$ on $S$ to a fixed point $x_0$ along a family of paths $p_{x_0, x(\sigma)}$. The point $x_0$ may lie on the surface $S$ or outside of it. In principle, the paths $p_{x_0, x(\sigma)}$ can be chosen freely, with different choices giving rise to different, inequivalent implementations of the parallel transported flux operator\footnote{Therefore, if one wished to use a fully explicit notation, the chosen family of paths should be included among the labels specifying the variable $\Et_i(S, x_0)$.}.

The key feature of the parallel transported flux variable is its simple behaviour under $SU(2)$ gauge transformations. Under a gauge transformation defined by a gauge function $g(x)\in SU(2)$, the variable \eqref{Et} transforms as
\begin{equation}
	\Et(S, x_0) \to g(x_0)\Et(S, x_0)g^{-1}(x_0).
	\label{g*Et}
\end{equation}

In order to derive the action of the parallel transported flux operator on a cylindrical function, it is useful to note the relation
\begin{equation}
	g\tau_i g^{-1} = D^{(1)}_{ki}(g)\tau_k
	\label{gtau}
\end{equation}
(which states that the generators $\tau_i$ transform under $SU(2)$ as the components of a vector). With the help of \Eq{gtau}, we can rewrite the classical variable \eqref{Et_i} as
\begin{equation}
	\Et_i(S, x_0) = \int_S d^2\sigma\,n_a(\sigma)D^{(1)}_{ki}\bigl(h_{x_0, x(\sigma)}^{-1}\bigr) E^a_k\bigl(x(\sigma)\bigr).
	\label{Et_i-2}
\end{equation}
When this expression is viewed as an operator and applied to a cylindrical function, the triad operator $E^a_k$ combined with the integral over the surface acts essentially as the standard flux operator, and compared to \Eq{Epsi} we simply pick up the rotation matrix $D^{(1)}_{ki}\bigl(h_{x_0, x(\sigma)}^{-1}\bigr)$ evaluated at the point where the surface intersects an edge. If we consider the action of the operator on a single holonomy, and assume that there is a single point of intersection $v$ between the edge and the surface, we obtain
\begin{equation}
	\Et_i(S, x_0)D^{(j)}(h_e) = D^{(1)}_{ki}\bigl(h_{x_0, v}^{-1}\bigr) E_k(S)D^{(j)}(h_e),
	\label{Et-action-1}
\end{equation}
where $E_k(S)$ is the regular flux operator acting on the holonomy. The result can be expressed in an alternative form by evaluating the action of the flux operator and then using \Eq{gtau} in reverse to eliminate the rotation matrix. For example, assuming for concreteness that $v$ is an interior point of the edge $e$, this calculation yields
\begin{equation}
	\Et_i(S, x_0)D^{(j)}(h_e) = i\nu(S, e)D^{(j)}(h_{e_2})D^{(j)}\bigl(h_{x_0, x_e}^{-1}\bigr)\tau_i^{(j)}D^{(j)}(h_{x_0, x_e})D^{(j)}(h_{e_1}).
	\label{Et-action-2}
\end{equation}
In practice, depending on the situation at hand, either one of \Eqs{Et-action-1} and \eqref{Et-action-2} may be the more convenient way of expressing the action of the parallel transported flux operator.

\subsection{The Hamiltonian constraint}
\label{sec:dynamics}

In the canonical formulation of loop quantum gravity, the dynamics of the theory is governed by the Hamiltonian constraint operator. Classically, the Hamiltonian constraint is given by the expression
\begin{equation}
	C = \frac{\epsilon\updown{ij}{k}E^a_iE^b_jF_{ab}^k}{\sqrt{\de}} - (1+\beta^2)\frac{E^a_iE^b_j}{\sqrt{\de}}\bigl(K_a^iK_b^j - K_a^jK_b^i\bigr)
	\label{C_orig}
\end{equation}
where $F_{ab}^i$ is the curvature of the Ashtekar--Barbero connection, $K_a^i$ is the extrinsic curvature of the spatial manifold, and $\beta$ is the Barbero--Immirzi parameter. In the case of vacuum gravity, the operator arising from \Eq{C_orig} is interpreted as a constraint operator, whose kernel defines the physical Hilbert space of the theory. Alternatively, one can consider a deparametrized formulation of gravity coupled with a scalar field, in which the scalar field is used as a relational time variable for the dynamics of the gravitational field \cite{quantized, HusainPawlowski, paper1, paper5}. In this case, the operator corresponding to \eqref{C_orig} (or a certain closely related operator) is interpreted as a physical Hamiltonian, which generates evolution with respect to the time variable provided by the scalar field.

The connection between the Hamiltonian constraint and the work presented in this article arises through the second term in \Eq{C_orig}, usually referred to as the Lorentzian part of the constraint. In loop quantum gravity, the traditional procedure of quantizing the Lorentzian term is due to Thiemann \cite{QSD}, and is based on a series of ingenious classical manipulations, as a result of which the second term of \Eq{C_orig} is expressed in terms of functions which correspond to well-defined operators in loop quantum gravity. More recently, there has emerged an alternative approach, which relies on the fact that (up to a term proportional to the Gauss constraint) the constraint \eqref{C_orig} can be rewritten in the form\footnote{Note the different numerical factor multiplying the Euclidean term in \Eq{C_new} relative to \Eq{C_orig}. The relevant classical identity states that the Lorentzian term of \Eq{C_orig} equals the curvature term of \Eq{C_new}, plus a multiple of the Euclidean term (plus a term proportional to the Gauss constraint).}
\begin{equation}
	C = \frac{1}{\beta^2}\frac{\epsilon\updown{ij}{k}E^a_iE^b_jF_{ab}^k}{\sqrt{\de}} + (1+\beta^2)\sqrt{\de}\Rt 
	\label{C_new}
\end{equation}
where $\Rt$ denotes the Ricci scalar of the spatial manifold. 

An operator representing the curvature part of the constraint \eqref{C_new} has been previously introduced in \cite{curvature}. The construction of \cite{curvature} makes use of the basic ideas of Regge calculus, and is based on the observation that the second term of \Eq{C_new}, integrated over the spatial manifold, happens to be the action integral of Euclidean gravity in three dimensions. Hence, this term can be expressed in terms of the hinge lengths and deficit angles associated to a cellular decomposition of a Regge-like, piecewise flat manifold approximating the smooth, physical spatial manifold $\Sigma$. Moreover, operators representing lengths and angles are readily available in loop quantum gravity, making it a relatively simple task to promote the Regge expression of the scalar curvature into a well-defined loop quantum gravity operator.

The operator which will be constructed in this article represents a new quantization of the curvature term in \Eq{C_new}, and therefore provides a novel approach towards defining the Hamiltonian constraint operator in loop quantum gravity. Since our quantization is based on writing the Ricci scalar directly as a function of the Ashtekar variables, the starting point of our construction is arguably more straightforward than the indirect approach of invoking Regge calculus to express the Ricci scalar in terms of quantizable objects. On the other hand, as already emphasized in the introduction, our construction does not encompass all states in the Hilbert space of loop quantum gravity, but is limited to states defined on a fixed graph forming a cubical lattice.

\section{Regularization of the Ricci scalar}

We now begin to move towards the main topic of this work, namely the construction of an operator representing the integrated Ricci scalar \eqref{int R}--\eqref{int NR}. In order to obtain such an operator, we must start by introducing a suitable regularization, as a result of which the integral is expressed in terms of objects -- \eg holonomies, fluxes, volumes -- which correspond to well-defined operators in loop quantum gravity.

The main technical challenge which must be dealt with in our construction is to find an appropriate regularization of the gauge covariant derivatives appearing in \Eq{qR}. At the moment we do not have a satisfactory proposal on how to accomplish this task for all states in the Hilbert space of loop quantum gravity, since these states may generally be based on graphs having a very complicated and irregular structure. In this work we will therefore restrict ourselves to considering the problem of defining the curvature operator on the Hilbert space of states based on a fixed cubical graph. (By a cubical graph we mean a graph whose nodes are six-valent, and whose edges are aligned with the coordinate directions defined by a fixed Cartesian background coordinate system.) However, as we will show in detail below, on the regular lattice provided by the cubical graph it becomes comparatively straightforward to regularize covariant derivatives by approximating them as finite differences between parallel transported flux variables associated to neighboring nodes of the graph. 

From the perspective of full loop quantum gravity, the assumption of a cubical graph appears to be a very significant limitation, since it implies that our construction applies only to a very small and specific sector of the entire Hilbert space of the theory. In the setting of the full theory, an operator defined on a fixed cubical graph can nevertheless be naturally interpreted within the framework of algebraic quantum gravity, which shows that a quantization of the full gravitational field can be achieved using just a single algebraic cubical graph. Moreover, as far as physical applications of loop quantum gravity are concerned, the restriction to a cubical graph does have sufficient generality to encompass a number of approaches attempting to probe the physical content of the theory. We have already mentioned quantum-reduced loop gravity and the effective dynamics program as two well-known examples of physical models which are formulated within the kinematical setting of full loop quantum gravity, and which make extensive use of states defined on cubical graphs. Hence our construction, as it stands, has a direct relevance to such models, and provides a well-defined curvature operator which is ready to be used in physical calculations in the context of these models.

\subsection{Overview of the regularization strategy}
\label{sec:strategy}

\begin{figure}[t]
\centering
	\includegraphics[scale=0.17]{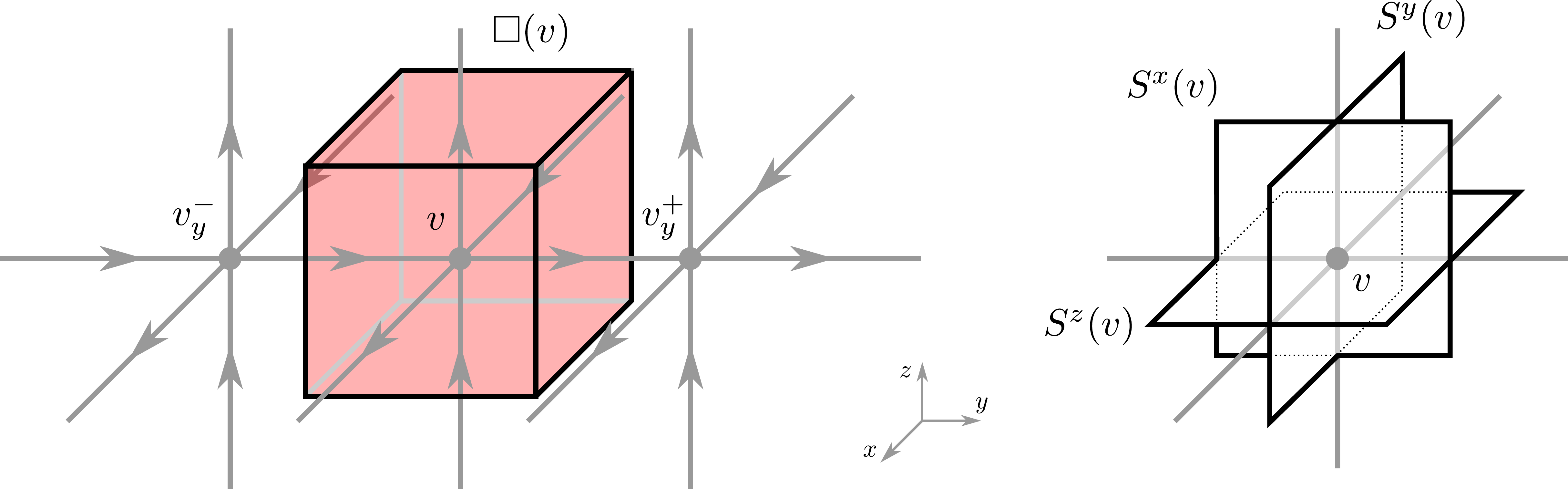}
	\caption{The structures involved in regularizing the integrated Ricci scalar. The integral $\int d^3x\,N\sqrt q\Rt$ is approximated as a discrete sum using the rectangular cells $\Box(v)$, each of which contains one node of the cubical spin network graph. Inside each cell $\Box(v)$ there are introduced three surfaces $S^a(v)$ ($a = x, y$ and $z$) which are dual to the corresponding background coordinate directions, and which are used to express the densitized triad and its derivatives in terms of flux variables.}
	\label{structures}
\end{figure}

We consider the regularization of the expression \eqref{int NR}, \ie the Ricci scalar integrated against an arbitrary smearing function. (The Ricci scalar \eqref{int R} itself can of course be recovered by setting $N = 1$ in the end.)

The structures involved in regularizing the integrated Ricci scalar are summarized in \Fig{structures}. In order to express the classical function defined by \Eqs{int NR} and \eqref{qR} in terms of variables which can be promoted into operators in loop quantum gravity, we introduce a partition of the spatial manifold into rectangular cells. Each cell contains a single node of the cubical spin network graph, and the faces of the cells are dual to the coordinate directions defined by the fixed background coordinate system. To make the presentation somewhat simpler, we assume that each cell is perfectly cubical, having coordinate volume $\epsilon^3$ and edge length $\epsilon$.

We let $\Box(v)$ denote the cell containing a particular node $v$ of the cubical graph. Within each cell $\Box(v)$ we set up three surfaces -- denoted by $S^x(v)$, $S^y(v)$ and $S^z(v)$ -- which are dual to the corresponding background coordinate directions, \ie the coordinate $x^a = {\rm const.}$ on the surface $S^a(v)$. Moreover, we require that the node $v$ is contained in each of the surfaces; in other words, the common point of intersection between all three surfaces is located at the node: $S^x(v)\cap S^y(v) \cap S^z(v) = v$. We also assume that the node $v$ coincides with the midpoint of the surface for all of the surfaces $S^a(v)$.

Now the starting point for regularizing the integrated Ricci scalar is to approximate the integral \eqref{int R} as a Riemann sum corresponding to the partition defined by the cells $\Box(v)$:
\begin{equation}
	\int d^3x\,N\sqrt q\Rt \simeq \sum_{\Box(v)} \epsilon^3 N(v)\sqrt{q(v)}\Rt(v).
	\label{sum R}
\end{equation}
Here $N(v)$, $q(v)$ and $\Rt(v)$ are the values of the smearing function, the metric determinant and the Ricci scalar at the point $v$. When \Eq{qR} is inserted for $\Rt(v)$ in \Eq{sum R}, the factor of $\epsilon^3$ can be distributed as follows between the various elements involved in the expression, with no factors of $\epsilon$ left over in the end:
\begin{itemize}
	\item $\epsilon^2$ for each factor of densitized triad,
	\item $\epsilon^{-2}$ for each inverse triad,
	\item $\epsilon^3$ for each factor of $\sqrt{\de}$,
	\item $\epsilon$ for each derivative (partial or covariant).
\end{itemize}
Therefore, each factor of densitized triad on the right-hand side of \Eq{sum R} can be regularized by replacing it with a flux variable associated to the corresponding surface $S^a(v)$, since at lowest order in the regularization parameter $\epsilon$ we have
\begin{equation}
	E_i\bigl(S^a(v)\bigr) \approx \epsilon^2E^a_i(v).
	\label{}
\end{equation}
In a similar way, the remaining kinds of factors entering the right-hand side of \Eq{sum R} -- covariant derivatives of the triad, as well as factors of inverse triad and the determinant of the triad, and their derivatives -- can be regularized by using appropriate combinations of holonomies and fluxes, and volumes of the cells $\Box(v)$. We will now consider the regularization of each of these elements in turn, starting with the covariant derivatives of the triad.

\subsection{Covariant derivatives of the triad}
\label{sec:derivatives}

The basic idea behind regularizing covariant derivatives of the densitized triad is to discretize them on the rectangular lattice provided by the cubical spin network graph, approximating a covariant derivative in terms of a finite difference of parallel transported flux variables located at neighboring nodes of the graph. For instance, a first derivative could be schematically discretized as
\begin{equation}
	f'(x) \simeq \frac{f(x+\epsilon) - f(x)}{\epsilon},
	\label{f'}
\end{equation}
where the points $x$ and $x+\epsilon$ are interpreted as two neighboring nodes of the graph. However, \Eq{f'} singles out the positive direction of the coordinate axis as the one which is used to perform the discretization. For our purposes this is not a very attractive feature, since if we are simply given a spin network state defined on a cubical graph, and consider a particular node $x$ and two of its neighboring nodes, there is no intrinsic way to determine which of them should be taken as the ''$x+\epsilon$'' that enters \Eq{f'}. In order to avoid introducing such an ambiguity, we prefer to discretize the derivatives in a way where the positive and negative coordinate directions are treated symmetrically. Thus, instead of \Eq{f'}, first derivatives will be regularized according to the symmetric discretization scheme
\begin{equation}
	f'(x) \simeq \frac{f(x+\epsilon) - f(x-\epsilon)}{2\epsilon},
	\label{f' sym}
\end{equation}
where $x+\epsilon$ and $x-\epsilon$ are to be understood as the two nodes immediately following and preceding the node $x$ in the direction of a given background coordinate axis. (This is the simplest possible discretization of a first derivative that does not invoke a preferred direction of the coordinate axis -- in other words, is invariant under the substitution $\epsilon\to -\epsilon$.)

In addition to first derivatives, we must also deal with the regularization of second covariant derivatives of the densitized triad. Here we have to consider separately the regularization of the ''pure'' second derivatives $\D_a^2E^b$ and the mixed second derivatives $\D_a\D_bE^c$, as the basic setup used for the regularization will be different in the two cases. For pure second derivatives, we take
\begin{equation}
	f''(x) \simeq \frac{f(x+\epsilon) - 2f(x) + f(x-\epsilon)}{\epsilon^2}
	\label{f''}
\end{equation}
as the basic pattern according to which the derivative is regularized. This expression, which represents the most straightforward discretization of the second derivative, is already symmetric between the positive and negative coordinate directions. For the mixed second derivatives of the triad, we will construct a suitable implementation of a symmetric discretization scheme
\begin{equation}
	\frac{\partial^2\!f(x,y)}{\partial x\partial y} \simeq \frac{f(x+\epsilon, y+\epsilon) - f(x+\epsilon, y-\epsilon) - f(x-\epsilon, y+\epsilon) + f(x-\epsilon, y-\epsilon)}{4\epsilon^2}.
	\label{f_xy}
\end{equation}
That is, the mixed second derivative $\D_a\D_bE^c$ at a given node $v$ will be regularized in terms of a discretization which uses the four nodes diagonally adjacent to the central node $v$ in the plane containing $v$ and spanned by the $x^a$- and $x^b$-coordinate axes of the background coordinate system.

Before proceeding to consider the detailed implementation of the construction outlined above, let us emphasize the reason why we base our construction on the use of gauge covariant derivatives and parallel transported flux variables, as opposed to partial derivatives and regular flux variables. In principle, it would be possible to construct a regularized expression for the integrated Ricci scalar starting from \Eq{qR_v1}, which expresses the Ricci scalar as a function of the densitized triad and its partial derivatives, and using finite differences of regular flux variables located at neighboring nodes of the cubical graph to discretize the partial derivatives. While such an expression would correctly approximate the integral of the Ricci scalar for small values of the regularization parameter, the difficulty with this approach is that it seems unclear how a gauge invariant\footnote{Under the internal $SU(2)$ gauge transformations defined by \Eq{gauge}.} curvature operator could be obtained upon quantization of the regularized expression. The basic problem is that the operators corresponding to the two terms in a finite difference of the form
\begin{equation}
	E_i\bigl(S^a(v')\bigr) - E_i\bigl(S^a(v)\bigr),
	\label{}
\end{equation}
where $v$ and $v'$ are two different nodes of the cubical graph, transform in different ways under internal gauge transformations, and hence the expression as a whole does not transform in any coherent manner. In contrast, as we will demonstrate below, by taking as our starting point the expression \eqref{qR} for the Ricci scalar in terms of the triad and its gauge covariant derivatives, we can use parallel transported flux variables to regularize the covariant derivatives in a way which will lead to a manifestly gauge invariant curvature operator.

\begin{figure}[p]
	\centering
	\includegraphics[scale=0.17]{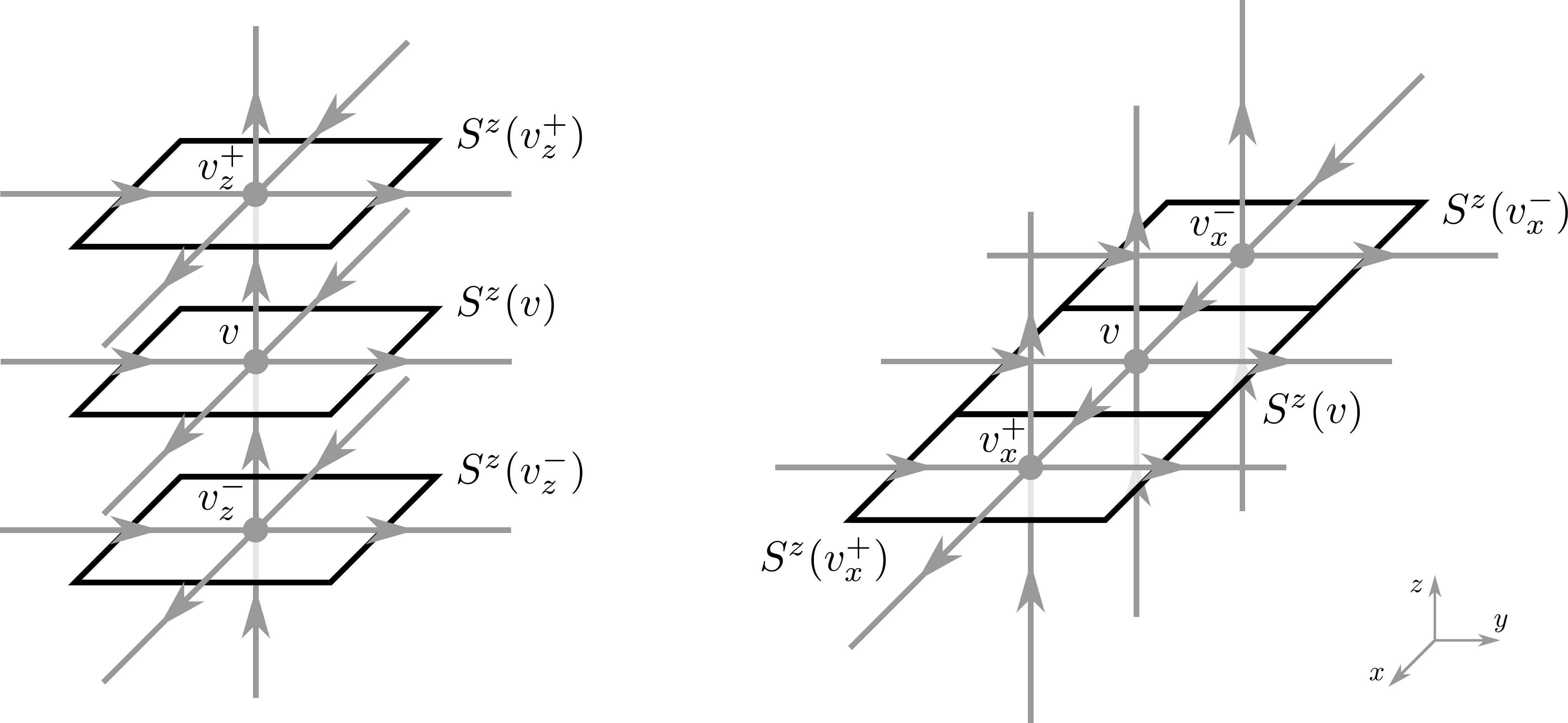}
	\caption{Regularization of gauge covariant derivatives of the densitized triad. Covariant derivatives of the triad at the node $v$ are regularized in terms of parallel transported flux variables associated to the surfaces shown in the figure. The parallel transport from a given node $v_a^\pm$ is taken to $v$ along the edge connecting the two nodes. The left diagram illustrates the regularization of the first derivative $\D_zE^z_i(v)$ and the second derivative $\D_z^2E^z_i(v)$, while the right diagram applies to the first derivative $\D_xE^z_i(v)$ and the second derivative $\D_x^2E^z_i(v)$.}
		\label{fig:dE}
\end{figure}

\begin{figure}[p]
	\centering
	\includegraphics[scale=0.17]{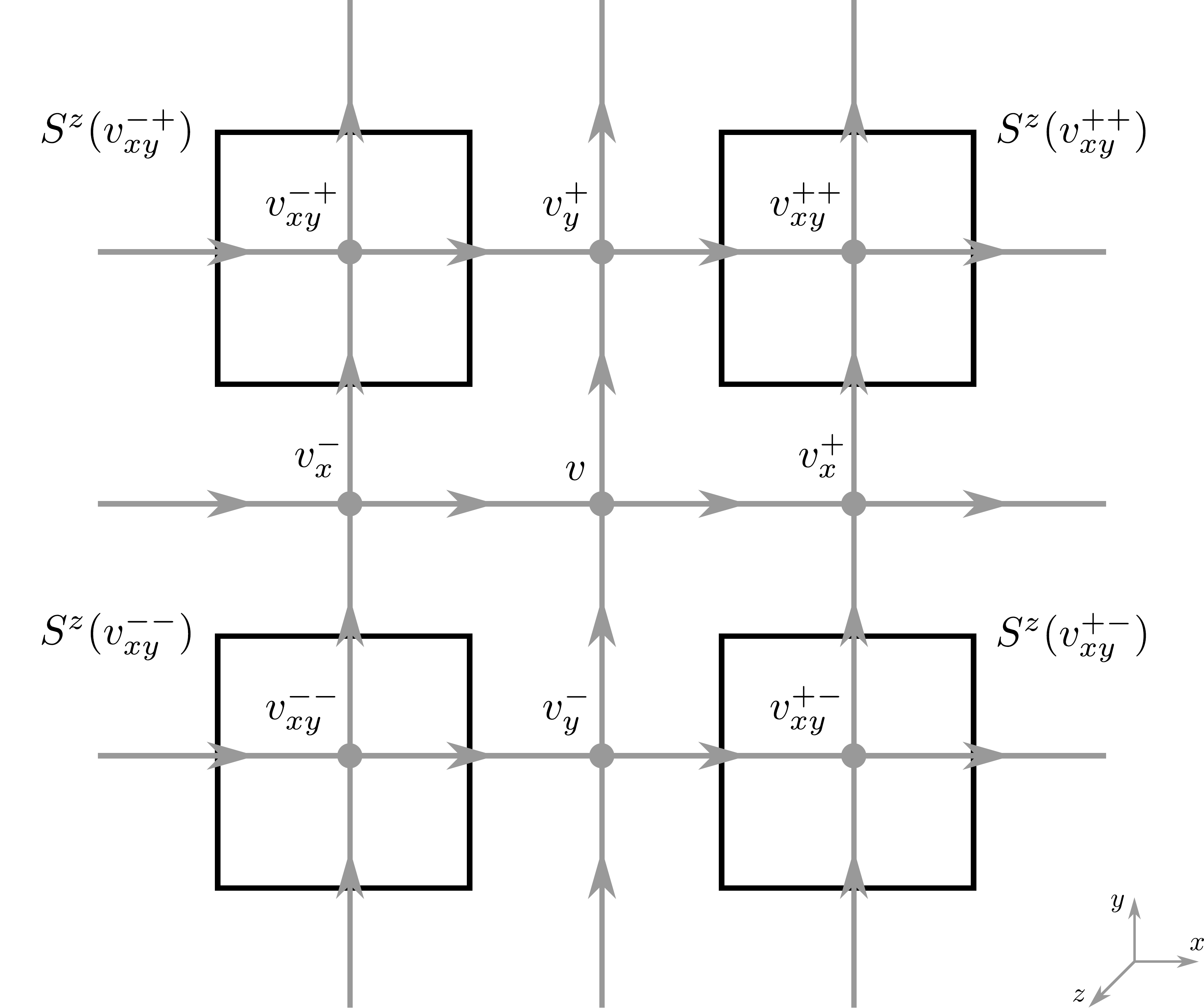}
	\caption{Regularization of the second covariant derivative $\D_x\D_yE^z_i$ at $v$. The four nodes diagonally adjacent to $v$ in the $xy$-coordinate plane are used to construct a regularized variable which approximates the symmetric part of the mixed second derivative.}
		\label{fig:ddE}
\end{figure}

\subsubsection*{First derivatives}

The setup for regularizing first covariant derivatives of the densitized triad is illustrated in \Fig{fig:dE}. Given a node $v$, we denote by $v_a^-$ and $v_a^+$ the nodes which come before and after $v$ in the direction of the background coordinate $x^a$. Following the symmetric discretization pattern indicated by \Eq{f' sym}, we then introduce
\begin{equation}
	\Delta_aE\bigl(S^b, v\bigr) \equiv \frac{\Et\bigl(S^b(v_a^+), v\bigr) - \Et\bigl(S^b(v_a^-), v\bigr)}{2}
	\label{DaE}
\end{equation}
as a regularized variable which approximates the covariant derivative $\D_aE^b$ at the point $v$. The flux variables $\Et\bigl(S^b(v_a^\pm), v\bigr)$ are parallel transported from their respective nodes to the central node $v$ as follows: The parallel transport to $v$ from a given point on the surface $S^b(v_a^\pm)$ is taken first to $v_a^\pm$ along a straight line lying within the surface, and then from $v_a^\pm$ to $v$ along the edge connecting the two nodes.

In order to verify that the expression \eqref{DaE} provides a valid regularization of the covariant derivative $\D_aE^b(v)$, we must check that it correctly approximates the covariant derivative for small values of the regularization parameter. The required calculations are performed in detail in Appendix \ref{sec:B}; here we will give just a brief outline. Letting $e_a^+$ and $e_a^-$ denote the edges which connect the nodes $v_a^+$ and $v_a^-$ to the central node $v$ (the orientation of the edges agreeing with the positive direction of the $x^a$-coordinate axis as shown in \Fig{fig:dE}), the parallel transported flux variables entering \Eq{DaE} can be written as
\begin{align}
	\Et\bigl(S^b(v_a^+), v\bigr) &= h_{e_a^+}^{-1}\Et\bigl(S^b(v_a^+), v_a^+\bigr)h_{e_a^+} \label{ES+} \\[1ex]
	\Et\bigl(S^b(v_a^-), v\bigr) &= h_{e_a^-}\Et\bigl(S^b(v_a^-), v_a^-\bigr)h_{e_a^-}^{-1} \label{ES-}
\end{align}
where the variables $\Et\bigl(S^b(v_a^\pm), v_a^\pm\bigr)$ are parallel transported to $v_a^\pm$ according to the straight-lines prescription specified in the text below \Eq{DaE}. Then one has to express all the holonomies and fluxes in terms of the Ashtekar variables $A_a$ and $E^a$, and expand the resulting expression in powers of the regularization parameter $\epsilon$. This calculation is presented in section \ref{DaE proof}, and yields the result
\begin{equation}
	\Et\bigl(S^b(v^a_\pm), v\bigr) = \epsilon^2E^b(v) \pm \epsilon^3\Bigl(\partial_aE^b(v) + \bigl[A_a(v), E^b(v)\bigr]\Bigr) + {\cal O}\bigl(\epsilon^4\bigr).
	\label{}
\end{equation}
Inserting this into \Eq{DaE}, and recognizing that the terms inside the parentheses are the covariant derivative $\D_aE^b(v)$, we find
\begin{equation}
	\Delta_aE\bigl(S^b, v\bigr) = \epsilon^3\D_aE^b(v) + {\cal O}\bigl(\epsilon^4\bigr),
	\label{DaE exp}
\end{equation}
showing that the expression \eqref{DaE} does give a correct regularization of the covariant derivative $\D_aE^b$ at the point $v$.

\subsubsection*{Second derivatives}

We continue to use the setup illustrated in \Fig{fig:dE} in order to construct a regularization for pure second derivatives of the triad. Following the pattern indicated by \Eq{f''}, and using again flux variables parallel transported to the central node $v$ (with the choice of paths from the surfaces to the node $v$ being the same as in the regularization of first derivatives), we introduce
\begin{equation}
	\Delta_{aa}E\bigl(S^b, v\bigr) \equiv \Et\bigl(S^b(v_a^+), v\bigr) - 2\Et\bigl(S^b(v), v\bigr) + \Et\bigl(S^b(v_a^-), v\bigr)
	\label{DaaEb}
\end{equation}
as a regularized variable approximating the second derivative $\D_a^2E^b$ at $v$. Regarding the variables entering the regularized expression, note that the only essential difference to the case of first derivatives is that here the central node $v$ is also involved the regularization, in addition to the nodes $v_a^+$ and $v_a^-$.

To verify the validity of \Eq{DaaEb} as a regularization of the derivative $\D_a^2E^b(v)$, we again have to show that the expression on the right-hand side reduces to the second covariant derivative at leading order in the parameter $\epsilon$. Essentially one has to repeat the steps that lead from \Eq{DaE} to \Eq{DaE exp}, except now all the variables must be expanded to one order higher in $\epsilon$. The calculation, which is performed in section \ref{DaaE proof}, leads in the end to the expected result 
\begin{equation}
	\Delta_{aa}E\bigl(S^b, v\bigr) = \epsilon^4\D_a^2E^b(v) + {\cal O}\bigl(\epsilon^5\bigr).
	\label{}
\end{equation}

Consider then the regularization of the mixed second derivatives $\D_a\D_bE^c(v)$ (with $a\neq b$). Keeping in mind the pattern given by \Eq{f_xy} for discretizing a mixed second derivative, we let $v_{ab}^{++}$, $v_{ab}^{+-}$, $v_{ab}^{-+}$ and $v_{ab}^{--}$ denote the four nodes diagonally adjacent to the central node $v$ in the plane which contains $v$ and is spanned by the $x^a$- and $x^b$-coordinate directions of the background coordinate system -- see \Fig{fig:ddE} for an illustration in the case of the derivative $\D_x\D_yE^z(v)$. The derivative $\D_a\D_bE^c(v)$ will be discretized in terms of flux variables associated to surfaces located at these four nodes and parallel transported to the central node. However, an issue which we encounter here is that there are two equally viable paths available for performing the parallel transport to $v$ from a given node along the edges of the cubical spin network graph. For example, the parallel transport from $v_{ab}^{++}$ to $v$ can be taken first from $v_{ab}^{++}$ to $v_a^+$ and then from $v_a^+$ to $v$, or first from $v_{ab}^{++}$ to $v_b^+$ and then to $v$.

The correct way to deal with this ambiguity is indicated by the fact that only the symmetric part of the second covariant derivative $\D_a\D_bE^c$ enters \Eq{qR}, which expresses the Ricci scalar as a function of the densitized triad and its covariant derivatives. If we construct a discretization of the derivative $\D_a\D_bE^c(v)$ by taking an average over the two possible paths every time a choice of path has to be specified, we will obtain an expression which is symmetric in $x$ and $y$. Thus, we expect that the regularized expression obtained in this way will approximate the symmetric part of the second covariant derivative, \ie $\D_{(a}\D_{b)}E^c(v)$.

To spell out the regularized expression corresponding to the idea described above, let us define the label $\sigma$ which takes the values $++$, $+-$, $-+$ and $--$, and hence labels the four nodes diagonally neighboring the central node. Furthermore, we introduce the formal vector
\begin{equation}
	\sigma^a = \bigl(\sigma^1, \sigma^2\bigr)
	\label{}
\end{equation}
whose components are equal to $+1$ or $-1$ according to the value of the label $\sigma$; for example, if $\sigma = ++$, then $\sigma^a = (1, 1)$. With this notation, we define
\begin{equation}
	\Et\bigl(S^c(v_{ab}^{\sigma}), v\bigr)_{\rm sym.} \equiv \frac{1}{2}\Bigl(\Et\bigl(S^c(v_{ab}^{\sigma}), v\bigr)_{v_{ab}^{\sigma} \to v_a^{\sigma^1} \to v} + \Et\bigl(S^c(v_{ab}^{\sigma}), v\bigr)_{v_{ab}^{\sigma} \to v_b^{\sigma^2} \to v}\Bigr)
	\label{}
\end{equation}
as a flux variable parallel transported from $v_{ab}^\sigma$ to $v$ symmetrically along the two available paths, the subscripts on the right-hand side indicating the path used for the parallel transport in each of the flux variables. Then, following the pattern of \Eq{f_xy}, we take
\begin{align}
	\Delta_{ab}E\bigl(S^c, v\bigr) \equiv \frac{1}{4}\biggl(\Et\bigl(&S^c(v_{ab}^{++}), v\bigr)_{\rm sym.} - \Et\bigl(S^c(v_{ab}^{+-}), v\bigr)_{\rm sym.} \notag \\
	&- \Et\bigl(S^c(v_{ab}^{-+}), v\bigr)_{\rm sym.} + \Et\bigl(S^c(v_{ab}^{--}), v\bigr)_{\rm sym.}\biggr)
	\label{DabEc}
\end{align}
as the regularized variable intended to approximate the symmetric part of the second covariant derivative $\D_a\D_bE^c$ at $v$. The calculation establishing the validity of the regularization is again given in Appendix \ref{sec:B}. In section \ref{DabE proof} we extract the leading term in the expansion of the expression \eqref{DabEc} in powers of $\epsilon$, and find
\begin{equation}
	\Delta_{ab}E\bigl(S^c, v\bigr) = \epsilon^4\D_{(a}\D_{b)}E^c(v) + {\cal O}\bigl(\epsilon^5\bigr),
	\label{}
\end{equation}
confirming that the variable \eqref{DabEc} indeed approximates the symmetric part of the second covariant derivative.

\subsection{Inverse triad and its derivatives}

In addition to the densitized triad itself and its covariant derivatives, we must also consider the regularization of the inverse triad
\begin{equation}
	E_a^i = \frac{1}{2\det E}\epsilon_{abc}\epsilon^{ijk}E^b_jE^c_k
	\label{E_a^i}
\end{equation}
as well as its first covariant derivatives. Note, however, that the expression \eqref{qR} has been arranged to not contain any second covariant derivatives of the inverse triad.

We begin by introducing the object
\begin{equation}
	\EEt_a^i\bigl(S(v), v'\bigr) = \frac{1}{2W(v)}\epsilon_{abc}\epsilon^{ijk}\Et_j\bigl(S^b(v), v'\bigr)\Et_k\bigl(S^c(v), v'\bigr)
	\label{EEt}
\end{equation}
where
\begin{equation}
	W(v) = \frac{1}{3!}\epsilon_{abc}\epsilon^{ijk}\Et_i\bigl(S^a(v), v\bigr)\Et_j\bigl(S^b(v), v\bigr)\Et_k\bigl(S^c(v), v\bigr)
	\label{W}
\end{equation}
(On the left-hand side of \Eq{EEt}, $S(v)$ is to be understood as a kind of a collective label for the two surfaces $S^b(v)$ $(b\neq a)$ involved in the flux variables on the right-hand side.) We then propose
\begin{equation}
	\EEt_a^i\bigl(S(v), v\bigr) = \frac{1}{2W(v)}\epsilon_{abc}\epsilon^{ijk}\Et_j\bigl(S^b(v), v\bigr)\Et_k\bigl(S^c(v), v\bigr)
	\label{E_a^i reg}
\end{equation}
as a regularized variable approximating the inverse triad $E_a^i$ at $v$. With the help of \Eq{E-square} from the appendix, it is immediate to see that
\begin{equation}
	W(v) = \epsilon^6\det E(v) + {\cal O}\bigl(\epsilon^8\bigr).
	\label{W=det}
\end{equation}
Using this and \Eq{E-square} itself in \Eq{E_a^i reg}, we find
\begin{equation}
	\EEt_a^i\bigl(S(v), v\bigr) = \epsilon^{-2}E_a^i(v) + {\cal O}\bigl(\epsilon^0\bigr),
	\label{}
\end{equation}
which shows that the regularized variable \eqref{E_a^i reg} indeed correctly approximates the inverse triad \eqref{E_a^i} at $v$, with the error term being of quadratic order in the regularization parameter $\epsilon$ relative to the leading term.

Consider then the variable
\begin{equation}
	\Delta_a\EEt_b^i(v) \equiv \frac{\EEt_b^i\bigl(S(v_a^+), v\bigr) - \EEt_b^i\bigl(S(v_a^-), v\bigr)}{2}
	\label{DEE}
\end{equation}
as a regularization of the covariant derivative $\D_aE_b^i(v)$. To verify that this proposal is correct, we recall from section \ref{sec:derivatives} that
\begin{equation}
	\Et_i\bigl(S^b(v_a^\pm), v\bigr) = \epsilon^2E^b_i(v) \pm \epsilon^3\D_aE^b_i(v) + {\cal O}\bigl(\epsilon^4\bigr).
	\label{L1}
\end{equation}
Moreover, \Eq{W=det} implies
\begin{align}
	\frac{1}{W(v^\pm)} = \epsilon^{-6}\frac{1}{\det E(v^\pm)} + {\cal O}\bigl(\epsilon^{-4}\bigr) = \epsilon^{-6}\frac{1}{\det E(v)} \pm \epsilon^{-5}\partial_x\biggl(\frac{1}{\det E}\biggr)\bigg|_v + {\cal O}\bigl(\epsilon^{-4}\bigr).
	\label{L2}
\end{align}
Combining \Eqs{L1} and \eqref{L2} with \Eq{EEt}, and using the fact that the covariant derivative obeys the Leibniz rule, a short calculation gives
\begin{equation}
	\EEt_b^i\bigl(S(v_a^\pm), v\bigr) = \epsilon^{-2}E_b^i(v) \pm \epsilon^{-1}\D_aE_b^i(v) + {\cal O}\bigl(\epsilon^0\bigr).
	\label{}
\end{equation}
From this it follows
\begin{equation}
	\Delta_a\EEt_b^i(v) = \epsilon^{-1}\D_aE_b^i(v) + {\cal O}\bigl(\epsilon^0\bigr),
	\label{}
\end{equation}
which confirms that the variable \eqref{DEE} does provide a valid regularization of the covariant derivative $\D_aE_b^i$ at $v$.

\subsection{Factors of $\de$}
\label{sec:det}

The factors of $\de$ appearing in \Eq{qR} can be regularized in terms of the volume of the cell $\Box(v)$,
\begin{equation}
	V(v) = \int_{\Box(v)} d^3x\,\sqrt q,
	\label{}
\end{equation}
which satisfies
\begin{equation}
	V(v) \approx \epsilon^3\sqrt{\dE v}
	\label{V approx}
\end{equation}
at leading order in the regularization parameter $\epsilon$. Since the volume is a gauge invariant observable (under the internal $SU(2)$ gauge transformations), the derivatives of $\de$ entering \Eq{qR} can be discretized as differences between volumes associated to neighboring nodes of the graph; there is no need to invoke any parallel transport operations in order to guarantee that the resulting expressions will be gauge invariant. Hence, the first derivative $\partial_a\de$ at $v$ is discretized as
\begin{equation}
	\Delta_a V(v)^2 = \frac{V(v_a^+)^2 - V(v_a^-)^2}{2},
	\label{ddet}
\end{equation}
the pure second derivative $\partial_a^2\de$ as
\begin{equation}
	\Delta_{aa} V(v)^2 = V(v_a^+)^2 - 2V(v)^2 + V(v_a^-)^2,
	\label{d2det}
\end{equation}
and the mixed second derivative $\partial_a\partial_b\de$ as
\begin{equation}
	\Delta_{ab} V(v)^2 = \frac{V(v_{ab}^{++})^2 - V(v_{ab}^{+-})^2 - V(v_{ab}^{-+})^2 + V(v_{ab}^{--})^2}{4}.
	\label{dddet}
\end{equation}

\subsection{The regularized Ricci scalar}

The regularization of the integrated Ricci scalar \eqref{int NR} can now be carried out using the elements introduced in the previous sections. In the regularized expression \eqref{sum R}, each factor of a densitized triad coming from \Eq{qR} is replaced with a flux through the corresponding surface $S^a(v)$. Similarly, each inverse triad is replaced with the regularized variable defined by \eqref{E_a^i reg} and each factor of $\sqrt{\de}$ is replaced with the volume $V(v)$, while all derivatives are replaced with the discretized derivatives defined in sections \ref{sec:derivatives}--\ref{sec:det}. As already mentioned in section \ref{sec:strategy}, no explicit dependence on the regularization parameter $\epsilon$ will remain after this procedure is carried out, reflecting the fact that the integrand in \Eq{int NR} carries a density weight of one.

For the sake of consistency, we will use parallel transported flux variables to regularize all instances of the densitized triad in \Eq{qR}, even those which are not acted on by derivatives. This is essentially a cosmetic modification, since it makes no difference to the form of the resulting quantum operator, but it does ensure that the regularized expression is invariant under $SU(2)$ gauge transformations already at the classical level. Accordingly, we define the regularized counterparts of the variables \eqref{Q^ab}, \eqref{Q_ab}, \eqref{Aabc} and \eqref{Babc} as
\begin{align}
	{\cal Q}^{ab}(v) &= \Et_i\bigl(S^a(v), v\bigr)\Et_i\bigl(S^b(v), v\bigr) \label{QQ^ab} \\
	{\cal Q}_{ab}(v) &= \widetilde{\cal E}_a^i\bigl(S(v), v\bigr)\widetilde{\cal E}_b^i\bigl(S(v), v\bigr) \label{QQ_ab} \\
	{\cal A}\updown{ab}{c}(v) &= \Et_i\bigl(S^a(v), v\bigr)\Delta_c E_i\bigl(S^b, v\bigr) \\
	{\cal B}\downup{ab}{c}(v) &= \widetilde{\cal E}_a^i\bigl(S(v), v\bigr)\Delta_b E_i\bigl(S^c, v\bigr).
	\label{}
\end{align} 
We have then managed to regularize the integrated Ricci scalar as
\begin{equation}
	\int d^3x\,N\sqrt q\Rt \simeq \sum_{\Box(v)} \frac{N(v)}{V(v)}{\cal R}(v)
	\label{int R reg}
\end{equation}
where
\begin{align}
	{\cal R}(v) = & -2\Et_i\bigl(S^a(v), v\bigr)\Delta_{ab}E_i\bigl(S^b, v\bigr) + 2{\cal Q}^{ab}(v)\widetilde{\cal E}_c^i\bigl(S(v), v\bigr)\Delta_{ab}E_i\bigl(S^c, v\bigr) \notag \\
	& -\Delta_aE_i\bigl(S^a, v\bigr)\Delta_bE_i\bigl(S^b, v\bigr) - \frac{1}{2}\Delta_aE_i\bigl(S^b, v\bigr)\Delta_bE_i\bigl(S^a, v\bigr) \notag \\
	& + \frac{5}{2}{\cal Q}^{ab}(v)\Delta_aE_i\bigl(S^c, v\bigr)\Delta_b{\cal E}_c^i(v) - \frac{1}{2}{\cal Q}^{ab}(v){\cal Q}_{cd}(v) \Delta_aE_i\bigl(S^c, v\bigr)\Delta_bE_i\bigl(S^d, v\bigr) \notag \\
	& + 2{\cal A}\updown{ab}{a}(v){\cal B}\downup{cb}{c}(v) + 2{\cal A}\updown{ab}{b}(v){\cal B}\downup{ca}{c}(v) + {\cal A}\updown{ab}{c}(v){\cal B}\downup{ba}{c}(v) \notag \\
	& + \frac{1}{2}{\cal Q}_{ab}(v){\cal A}\updown{ca}{d}(v){\cal A}\updown{db}{c}(v) - {\cal Q}^{ab}(v){\cal B}\downup{ca}{c}(v){\cal B}\downup{db}{d}(v) \notag \\
	& + 2\bigl({\cal Q}^{ab}(v){\cal B}\downup{ca}{c}(v) - {\cal A}\updown{ab}{a}(v) - {\cal A}\updown{ba}{a}(v)\bigr)\frac{\Delta_b V(v)^2}{V(v)^2} \notag \\
	& + \frac{3}{2}{\cal Q}^{ab}(v)\frac{\Delta_a V(v)^2}{V(v)^2}\frac{\Delta_b V(v)^2}{V(v)^2} - 2{\cal Q}^{ab}(v)\frac{\Delta_{ab}V(v)^2}{V(v)^2}.
	\label{R(v)}
\end{align}
For small values of the regularization parameter $\epsilon$, the regularized expression on the right-hand side of \Eq{int R reg} converges to the continuum expression on the left-hand side.

\section{The curvature operator}

\subsection{Quantization of the regularized Ricci scalar}

By carrying out the regularization detailed in the previous chapter, we have expressed the integrated Ricci scalar in a form suitable for quantization. Every factor involved in the expression \eqref{R(v)} can now be readily promoted into a well-defined operator of loop quantum gravity. Each instance of a parallel transported flux variable is naturally replaced with the corresponding parallel transported flux operator. Moreover, each appearance of the volume $V(v)$ is replaced with the volume operator
\begin{equation}
	V_v = \sqrt{|q_v|}
	\label{}
\end{equation}
acting on the node $v$. For the six-valent nodes of the cubical graph, the operator $q_v$ entering the definition of the volume operator can be expressed in terms of the flux operators $\Et_i\bigl(S^a(v), v\bigr)$ as
\begin{equation}
	q_v = \epsilon^{ijk}\Et_i\bigl(S^x(v), v\bigr)\Et_j\bigl(S^y(v), v\bigr)\Et_k\bigl(S^z(v), v\bigr).
	\label{}
\end{equation}
The negative powers of the volume $V(v)$ are quantized in terms of the regularized inverse volume operator ${\cal V}_v^{-1}$, which is defined by specifying its spectral decomposition as follows.\footnote{
	The operator ${\cal V}_v^{-1}$ was first used in loop quantum gravity by Bianchi in his construction of the length operator \cite{Bianchi}. It has since been used \eg in the quantization of the scalar curvature based on Regge's formula \cite{curvature}, and in the construction of a Hamiltonian constraint operator playing the role of the physical Hamiltonian in a model of loop quantum gravity deparametrized with respect to non-rotational dust \cite{paper5, paper2}.

More generally, one could consider the operator ${\cal V}_v^{-1}(\delta) \equiv V_v/(V_v^2 + \delta^2)$ -- the so-called Tikhonov regularization of the inverse operator -- where the parameter $\delta$ has a small but finite value. The operator defined by \Eq{V_v^-1} can be obtained by taking the limit $\delta\to 0$ of the operator ${\cal V}_v^{-1}(\delta)$.}
Let $\ket\lambda$ be an eigenstate of $V_v$, the standard volume operator restricted to the node $v$, with eigenvalue $\lambda$. Then the action of ${\cal V}_v^{-1}$ on the state $\ket\lambda$ is defined to be
\begin{equation}
	{\cal V}^{-1}_v\ket\lambda = \begin{cases} 
		\lambda^{-1}\ket\lambda & \text{if $\lambda\neq 0$} \\[0.5ex] 
		0 & \text{if $\lambda = 0$}
	\end{cases}
	\label{V_v^-1}
\end{equation}
The factor of $1/W(v)$, which appears in the regularized inverse triad \eqref{E_a^i reg}, can be dealt with in the same way. The squared, oriented volume $W(v)$ corresponds to the operator $q_v$. Hence, if $\ket\mu$ is an eigenstate of $q_v$ with eigenvalue $\mu$, we define the regularized inverse operator corresponding to $1/W(v)$ as
\begin{equation}
	{\cal W}^{-1}_v\ket\mu = \begin{cases}
		\mu^{-1}\ket\mu & \text{if $\mu\neq 0$} \\[0.5ex]
		0 & \text{if $\mu=0$}
	\end{cases}
	\label{W_v^-1}
\end{equation}
In this way we have obtained an operator representing the Ricci scalar integrated against a smearing function. The action of the operator on a cylindrical function based on the cubical graph $\Gc$ takes the form
\begin{equation}
	\biggl(\widehat{\int d^3x\,N\sqrt q\Rt}\biggr)\ket{\Psi_\Gc} = \sum_{v\in{\Gc}} N(v){\cal V}_v^{-1}{\cal R}_v\,\ket{\Psi_\Gc}
	\label{NR-op}
\end{equation}
where ${\cal R}_v$ is the operator corresponding to the expression \eqref{R(v)}. (An operator representing the actual scalar curvature of the spatial manifold, \ie $\int d^3x\sqrt q\Rt$, can of course be recovered simply by setting $N = 1$ in the above expression.)

\subsection{Discussion}
\label{properties}

\subsubsection*{Gauge and diffeomorphism invariance}

The operator \eqref{NR-op} is invariant under the internal $SU(2)$ gauge transformations generated by the Gauss constraint. This property is accomplished by our choice to start the construction of the operator by expressing the classical Ricci scalar in terms of gauge covariant derivatives of the triad, and to regularize these derivatives in terms of parallel transported flux variables which are all transported to the same node of the graph. In the framework of full loop quantum gravity, an operator representing a geometrical observable, such as the integrated Ricci scalar \eqref{int R}, should additionally be invariant under spatial diffeomorphisms, while the Ricci scalar integrated against an arbitrary smearing function should be represented by an operator transforming covariantly under diffeomorphisms. However, our choice to define the curvature operator on the Hilbert space of a single fixed graph excludes the possibility of directly studying the transformation properties of the operator under diffeomorphisms (\eg the group of diffeomorphisms which map cubical graphs to other cubical graphs), although the operator is trivially invariant under diffeomorphisms which preserve the chosen cubical graph.

In the context of an operator defined on a fixed graph, the most natural set of tools for discussing diffeomorphism invariance appears to be that provided by algebraic quantum gravity. After all, there one is forced to develop methods of addressing the diffeomorphism constraint in a way which is consistent with the choice to work with a single graph only. One possibility would be to include the diffeomorphism constraint as a part of the so-called extended master constraint \cite{master1, master2}, whose role is to select the gauge invariant states and observables of the theory. Another option is the reduced phase space quantization of \cite{aqg4}, where a family of four Brown--Kucha\v{r} scalar fields is used to deparametrize the entire spacetime manifold, after which the diffeomorphism constraint essentially disappears and the physical Hilbert space is formed by all $SU(2)$-invariant states. Regardless of which approach is chosen, the operator constructed in this article could play a role in its implementation, entering the construction of the master constraint operator or the physical Hamiltonian governing the dynamics of the deparametrized theory.

\subsubsection*{Quantization ambiguities}

The construction leading up to the operator defined by \Eqs{NR-op} and \eqref{R(v)} involves several quantization ambiguities, and consequently the operator obtained as the result of the construction is far from being uniquely determined. (Of course, such ambiguities are routinely encountered in loop quantum gravity, and are typically found in any operator representing a moderately complicated classical function, so their presence here is not particularly alarming.)

In general, the operators corresponding to the various factors in \Eq{R(v)} do not commute with each other, so there exist many possible, inequivalent factor orderings of the operator \eqref{NR-op}. Another source of ambiguities arises from the fact that one can use the identities $E_a^i\D_bE^a_j = -E^a_j\D_bE_a^i$ and $E_a^i\D_bE^c_i = -E^c_i\D_bE_a^i$ to rewrite the classical expression \eqref{qR} in multiple different ways which are all equal to each other classically but lead to inequivalent operators upon quantization. Furthermore, the discretization schemes used to regularize covariant derivatives are certainly not unique, although the choices we have made are distinguished by being the simplest possible ones that respect the requirement of symmetry between positive and negative directions of the background coordinate axes. However, from the point of view of simply obtaining a mathematically well-defined operator, nothing would prevent one from using a more complicated discretization, which could involve nodes more distant from the central node, or a higher number of nodes overall, or in which the parallel transport to the central node is taken along a path of more complicated shape (which may in principle contain segments pointing along every possible coordinate direction). After having selected a discretization scheme, one should only repeat the calculations performed in Appendix \ref{sec:B} in order to verify whether the chosen regularized variables correctly approximate the covariant derivatives of the triad in the limit of small discretization parameter.

At a first sight, it might seem that the spin carried by the holonomies entering the parallel transported flux operator is also subject to an ambiguity. After all, a similar situation arises in the construction of the Hamiltonian constraint operator, where the holonomies involved in the operator can be regularized using any irreducible representation of $SU(2)$. However, in our case it is straightforward to verify that all representations are equivalent for the purpose of defining the parallel transported flux operator. Instead of \Eqs{Et_i} and \eqref{Et} where the fundamental representation is used, one could use the spin-$j$ representation to define the parallel transported flux variable by the equations
\begin{equation}
	\Et_i^{(j)}(S, x_0) = -\frac{1}{N_j}\,{\rm Tr}\,\Bigl(\tau_i^{(j)} \Et^{(j)}(S, x_0)\Bigr)
	\label{Et_i^j}
\end{equation}
\begin{equation}
	\Et^{(j)}(S, x_0) = \int_S d^2\sigma\,n_a(\sigma)D^{(j)}\bigl(h_{x_0, x(\sigma)}\bigr)E^a_i\bigl(x(\sigma)\bigr)\tau_i^{(j)}D^{(j)}\bigl(h^{-1}_{x_0, x(\sigma)}\bigr)
	\label{Et^j}
\end{equation}
where $N_j = j(j+1)(2j+1)/3$, and the numerical factor is determined by the normalization ${\rm Tr}\,\bigl(\tau_i^{(j)}\tau_k^{(j)}\bigr) = -N_j\delta_{ik}$ of the $SU(2)$ generators. However, in any irreducible representation there holds the relation
\begin{equation}
	D^{(j)}(g)\tau_i^{(j)}D^{(j)}(g^{-1}) = D^{(1)}_{ki}(g)\tau_k^{(j)}
	\label{}
\end{equation}
with the spin-$1$ representation matrix acting on the vector index of the generator on the right-hand side. It follows that the variable defined by \Eqs{Et_i^j} and \eqref{Et^j} is actually independent of $j$, and for any value of $j$ is equal to
\begin{equation}
	\Et_i^{(j)}(S, x_0) = \int_S d^2\sigma\,n_a(\sigma)D^{(1)}_{ki}\bigl(h_{x_0, x(\sigma)}^{-1}\bigr) E^a_k\bigl(x(\sigma)\bigr).
	\label{}
\end{equation}
This implies that the spin carried by the holonomy operators arising from the parallel transported flux is not subject to a choice, but is fixed to be equal to $1$. Thus, the parallel transported flux operators involved in the curvature operator will act by coupling holonomies of spin $1$ with the holonomies present in the state on which we apply the operator, regardless of which spin was originally used to define the parallel transported flux operator.

\subsubsection*{The adjoint operator}

The operator ${\cal R}_v$, being defined on the Hilbert space of a fixed graph, possesses a densely defined adjoint operator ${\cal R}_v^\dagger$ on this space. Using the operators ${\cal R}_v$ and ${\cal R}_v^\dagger$, we may then introduce a symmetric factor ordering of the operator \eqref{NR-op}, for instance as
\begin{equation}
\sum_{v\in\Gamma_0} N(v){\cal V}_v^{-1/2}\frac{{\cal R}_v + {\cal R}_v^\dagger}{2}{\cal V}_v^{-1/2}.
	\label{}
\end{equation}
The possibility of defining a symmetric factor ordering is a necessary requirement from the perspective that any classically real-valued geometrical observable, such as the scalar curvature, should be represented by a self-adjoint quantum operator. In addition, the symmetric form of the curvature operator can be used as a part of the physical Hamiltonian in models of loop quantum gravity deparametrized with respect to a scalar field, where the Hamiltonian is interpreted as the generator of physical time evolution, and therefore has to be a self-adjoint operator.

We may note that if we tried to extend our construction, in its present form, to define a curvature operator on the entire Hilbert space of loop quantum gravity (which includes states based on all possible graphs), we would encounter a known problem which would prevent the adjoint operator from being densely defined. The issue arises from the fact that the holonomies contained in the parallel transported flux operators act by changing the spin quantum numbers of the state on which the curvature operator acts, and in some cases the action of the operator produces a state where the spin of an edge has become equal to zero -- in other words, the edge has been ''erased'' from the graph. On the full Hilbert space the action of the adjoint operator on such a state will be ill-defined, essentially because there are infinitely many inequivalent ways in which the adjoint operator can reintroduce the missing edge (even if we are working at the level of diffeomorphism invariant states).\footnote{For a more detailed discussion of this point, see \eg section 13.5 of \cite{thesis}, where the phenomenon of the ''disappearing edge'' is examined in the context of Thiemann's regularization of the Hamiltonian constraint.} For this reason, it is necessary to introduce some limitation on the set of graphs being considered -- \eg by restricting to just a single fixed graph, as we have chosen to do -- in order to ensure that a symmetric operator can be obtained as the result of the construction.

\subsubsection*{Possible alternative definitions}

Instead of defining the curvature operator on the Hilbert space of a fixed cubical graph, one may ask whether the operator could be defined instead on a space which would be analogous to the diffeomorphism invariant Hilbert space of full loop quantum gravity, but where the averaging is performed only with respect to diffeomorphisms preserving the cubical structure of the graph\footnote{Viewed as coordinate transformations on the spatial manifold, such diffeomorphisms can be characterized by their having the form
\begin{equation}
	(x, y, z) \to \bigl(X(x), Y(y), Z(z)\bigr)
\label{}
\end{equation}
where each new coordinate is a function of the corresponding old coordinate only.}. However, the action of our curvature operator, as we have defined in this article, is sensitive to the difference between the lack of an edge and the presence of an edge carrying spin zero. Consequently, it is not difficult to find examples where the operator acts in different ways on different representatives of the same diffeomorphism equivalence class, and so the definition of the operator cannot be consistently extended in the way envisioned above. Let us therefore briefly consider some possibilities of modifying the construction in such a way that the resulting operator behaves consistently under diffeomorphisms on the cubical lattice, and whose definition could therefore be extended to the space of states invariant under such diffeomorphisms.

One possible modification would consist of manually adjusting the definition of the discretized derivative operators so that they no longer act on edges carrying spin zero. For instance, the action of the operator $\Delta_aE_i(S^b, v)$ would be given by
\begin{equation}
	\Delta_aE_i\bigl(S^b, v\bigr) = \frac{\Et_i\bigl(S^b(v_a^+), v\bigr) - \Et_i\bigl(S^v(v_a^-), v\bigr)}{2}
	\label{}
\end{equation}
only on states where the edges connecting $v_a^+$ and $v_a^-$ to $v$ both carry a nonzero spin, but is declared to vanish if the spin on either of these edges is equal to zero. By modifying all the derivative operators in this way, one would obtain an operator which respects the notion of cylindrical consistency on the cubical lattice. However, the adjoint of this operator would act in a highly non-graph preserving manner, even if it is applied to a generic state in which every edge of the cubical graph carries a non-vanishing spin. From a practical perspective, such an operator would be unnecessarily complicated to work with, even if it would be a perfectly well-defined operator from the mathematical point of view.

Another possible way of modifying the definition of the curvature operator is to introduce projection operators, as proposed \eg in \cite{aqg4} and \cite{Thiemann2020}, in order to ensure that the operator acts in a strictly graph-preserving manner, so that its action neither creates any new edges nor destroys any edges originally present in the state on which the operator is acting. For example, if $\Gamma$ is any subgraph of the chosen cubical graph $\Gamma_0$, let $\widetilde{\cal H}_\Gamma$ denote the space spanned by spin network states on $\Gamma$ carrying a nonzero spin on every edge of the graph, and $P_\Gamma$ the orthogonal projection onto $\widetilde{\cal H}_\Gamma$. Instead of
\begin{equation}
	{\cal R} = \sum_{v\in\Gamma_0} {\cal V}_v^{-1/2}{\cal R}_v{\cal V}_v^{-1/2}
	\label{}
\end{equation}
one would then define the curvature operator by the expression
\begin{equation}
	\widetilde{\cal R} = \sum_{\Gamma\subset\Gamma_0} P_\Gamma{\cal R}P_\Gamma
\label{tilde R}
\end{equation}
where the sum runs over all subgraphs of $\Gamma_0$. The operator defined by \Eq{tilde R} transforms consistently under diffeomorphisms on the cubical lattice. Moreover, since the projection operators guarantee that the action of the operator is strictly graph-preserving, the same property will hold for the adjoint operator as well, which can be seen as a definite advantage over the first proposal discussed above.

\section{Conclusions}

In this article we have introduced a new geometric operator representing the scalar curvature of the three-dimensional spatial manifold in loop quantum gravity. While our operator is constructed using the basic kinematical structures of loop quantum gravity, it is not defined on the entire kinematical Hilbert space of the theory, but only on the Hilbert space of a fixed cubical graph (the graph being defined with respect to a fiducial Cartesian background coordinate system). In the context of full loop quantum gravity, perhaps the most natural framework for interpreting an operator of this type is provided by algebraic quantum gravity, which uses the mathematical apparatus of loop quantum gravity to perform a quantization of the full gravitational field entirely in terms of a single (abstract) cubical graph.

The starting point of our construction of the curvature operator is to express the classical Ricci scalar directly as a function of the Ashtekar variables. More specifically, the Ricci scalar is expressed in terms of the densitized triad and its $SU(2)$-covariant derivatives. The resulting expression must then be regularized by writing it in terms of objects which correspond to well-defined operators in loop quantum gravity. From the technical point of view, the main challenge at this step consists of constructing a suitable regularization of the gauge covariant derivatives of the triad. Having restricted ourselves to working on a cubical graph, it becomes a relatively simple task to regularize the covariant derivatives by discretizing them on the lattice provided by the graph in terms of finite differences of parallel transported flux variables. The use of gauge covariant derivatives and parallel transported flux variables guarantees that, as a geometrical observable, the operator representing the Ricci scalar is invariant under the internal $SU(2)$ gauge transformations generated by the Gauss constraint.

In addition to algebraic quantum gravity, the operator introduced in this article is relevant to various physical models of loop quantum gravity, which are formulated in terms of states defined on cubical graphs. Well-known examples of such approaches include quantum-reduced loop gravity, and models based on effective Hamiltonians derived from semiclassical states. For models of this type, our construction provides a well-defined curvature operator which can be used in physical applications. Moreover, when it comes to the physical properties of the operator, we expect that our construction represents an improvement over the earlier work in \cite{curvature}, where the basic notions of Regge calculus are invoked to define a scalar curvature operator for loop quantum gravity. In the companion article \cite{part2} we consider the operator introduced in this article in the framework of quantum-reduced loop gravity, and find that it gives rise to a non-trivial and seemingly satisfactory curvature operator on the Hilbert space of the quantum-reduced model, in contrast to the operator of \cite{curvature}, whose action gives trivially zero on any state in the quantum-reduced Hilbert space.

From the perspective of full loop quantum gravity, the restriction to a cubical graph certainly represents a rather serious limitation, since it implies that our curvature operator is defined only in a small subspace of the entire kinematical Hilbert space of the theory. However, in addition to facilitating the regularization of derivatives of the triad, the assumption of a cubical graph fulfills another important role, having to do with the requirement that any geometrical observable -- such as the scalar curvature -- should be represented in the quantum theory by a self-adjoint operator. Namely, the assumption that our operator is defined on the Hilbert space of a fixed graph ensures that its adjoint is available as a densely defined operator, and hence a symmetric factor ordering of the operator can be prescribed.

Note that the above statement applies to any operator involving holonomies associated to edges which overlap, even partially, with the edges of the graph on which the operator is acting. In general, the adjoint of such an operator cannot be densely defined on the entire kinematical Hilbert space of loop quantum gravity. Thus it seems certain that the regularization of derivatives in terms of parallel transported flux variables would have to be modified in some suitable way, if one would eventually like to refine the construction performed in this article into an operator which is well-defined in the proper framework of full loop quantum gravity, as opposed to being restricted to the Hilbert space of a single fixed graph.

To sketch a concrete idea of how such a modification could be achieved, we recall the work presented in \cite{paper1}, where a new regularization of the Euclidean part of the Hamiltonian is proposed. Unlike Thiemann's original regularization of the Hamiltonian \cite{QSD}, the adjoint of the operator introduced in \cite{paper1} is densely defined on the full kinematical Hilbert space of loop quantum gravity. The key idea behind the construction is that the loops created by the operator do not overlap with any edges of the graph on which the operator acts; instead, each loop is tangent to the pair of edges to which it is associated. Adapting this concept to the regularization of the curvature, we would be lead to consider parallel transported fluxes where the parallel transport from one node to another is not taken along the edge already present in the graph, but instead along a new edge connecting the two nodes (and being tangent to the previously existing edges at appropriate orders of tangentiality).  We leave the detailed investigation of an operator of this type as a possible topic for future work. \\

\begin{center}
\large{\bf{Acknowledgments}}
\end{center}

\noindent We are grateful to Cong Zhang for pointing out an error in the first version of the manuscript. \xspace{I. M.} thanks Mehdi Assanioussi for helpful discussions. This work was supported by grant \xspace{no.} 2018/30/Q/ST2/00811 of the Polish National Science Center.

\appendix

\section{Ricci scalar as a function of the densitized triad}
\label{sec:A}

In this Appendix we present a derivation of \Eq{qR}, which expresses the Ricci scalar of the spatial manifold in terms of the densitized triad and its gauge covariant derivatives. We begin by obtaining an expression for the Ricci scalar as a function of the triad and its partial derivatives, after which the desired result is found by writing the partial derivatives in terms of their gauge covariant counterparts.

The Ricci scalar is given by
\begin{equation}
	\Rt = q^{ab}\bigl(\partial_c\Gamma^c_{ab} - \partial_b\Gamma^c_{ac} + \Gamma^c_{ab}\Gamma^d_{cd} - \Gamma^c_{ad}\Gamma^d_{bc}\bigr)
	\label{R3}
\end{equation}
where the Christoffel symbols are
\begin{equation}
	\Gamma^a_{bc} = \frac{1}{2}q^{ad}\bigl(\partial_bq_{dc} + \partial_cq_{bd} - \partial_dq_{bc}\bigr).
	\label{Gamma}
\end{equation}
The spatial metric and its inverse are related to the densitized triad as
\begin{equation}
	q_{ab} = \de E_a^iE_b^i, \qquad q^{ab} = \frac{E^a_iE^b_i}{\de}
	\label{qq}
\end{equation}
where $\det E \equiv \det E^a_i$, and
\begin{equation}
	E_a^i = \frac{1}{2\det E}\epsilon_{abc}\epsilon^{ijk}E^b_jE^c_k
	\label{}
\end{equation}
is the inverse of the densitized triad, \ie it satisfies
\begin{equation}
	E^a_iE^i_b = \delta^a_b, \qquad E^a_iE^j_a = \delta_i^j.
	\label{}
\end{equation}
Now the required calculation consists simply of inserting \Eqs{Gamma} and \eqref{qq} into \Eq{R3}. In order to make the resulting expressions somewhat more compact, we make use of the following abbreviations for various combinations of the triad and its derivatives:

\begin{align}
	Q^{ab} &= E^a_iE^b_i \\
	Q_{ab} &= E_a^iE_b^i \\
	A\updown{ab}{c} &= E^a_i\partial_cE^b_i \\
	B\downup{ab}{c} &= E_a^i\partial_bE^c_i \\
	C\updown{a}{bc} &= E^a_i\partial_bE_c^i \\
	C_{abc} &= E^i_a\partial_bE_c^i \\
	S_{abcd} &= E_a^i\partial_b\partial_cE_d^i \\
	S\updown{a}{bcd} &= E^a_i\partial_b\partial_cE_d^i \\ 
	T_{abcd} &= (\partial_aE_b^i)(\partial_cE_d^i) \\
	U{}\updown{a}{bcd} &= (\partial_bE^a_i)(\partial_cE_d^i) \\
	L_a &= \partial_a\ln\de \\
	L_{ab} &= \partial_a\partial_b\ln\de
\end{align}
Note that most of the objects defined above are not tensors, which should be kept in mind when manipulating them.

\subsection{Christoffel symbols and their derivatives}

We start by using \Eq{qq} in \Eq{Gamma} to write the Christoffel symbols in terms of the densitized triad. A simple calculation yields
\begin{equation}
	\;\;\; \Gamma^a_{bc} = \frac{1}{2}\Bigr(C\updown{a}{bc} + C\updown{a}{cb} + Q^{ad}\bigl(C_{bcd} + C_{cbd} - C_{bdc} - C_{cdb}\bigr) + \delta^a_bL_c + \delta^a_cL_b  - Q^{ad}Q_{bc}L_d\Bigr). \;\;\;
	\label{G(E)}
\end{equation}
To evaluate the derivative of the Christoffel symbol, it is useful to first establish the identities
\begin{align}
	\partial_dC\updown{a}{bc} &= U{}\updown{a}{dbc} + S\updown{a}{dbc}, \\
	\partial_dC_{abc} &= T_{dabc} + S_{adbc}.
	\label{dG(E)}
\end{align}
Using these in \Eq{G(E)}, we immediately find
\begin{align}
	\partial_d\Gamma^a_{bc} = \frac{1}{2}\Bigr(&U{}\updown{a}{dbc} + U{}\updown{a}{dcb} + S\updown{a}{dbc} + S\updown{a}{dcb} + \partial_dQ^{ae}\bigl(C_{bce} + C_{cbe} - C_{bec} - C_{ceb}\bigr) \notag \\
	&+ Q^{ae}\bigl(T_{dbce} + T_{dcbe} - T_{dbec} - T_{dceb} + S_{bcde} + S_{cbde} - S_{bdec} - S_{cdeb}\bigr) \notag \\
&+ \delta^a_bL_{cd} + \delta^a_cL_{bd} - Q^{ae}(\partial_dQ_{bc})L_e - Q_{bc}(\partial_dQ^{ae})L_e - Q^{ae}Q_{bc}L_{de}\Bigr).
	\label{}
\end{align}

\subsection{Ricci scalar. Terms with derivatives of Christoffel symbols}

We now apply \Eqs{G(E)} and \eqref{dG(E)} to \Eq{R3} in order to find an expression for the Ricci scalar as a function of the densitized triad. We consider the four terms in \Eq{R3} one by one. The first term amounts to calculating
\begin{align}
	&E^b_iE^c_i\partial_a\Gamma^a_{bc} \notag \\
	&= Q^{bc}\biggl(U{}\updown{a}{abc} + S\updown{a}{abc} + \partial_aQ^{ae}\bigl(C_{bce} - C_{bec}\bigr) + Q^{ae}\bigl(T_{abce} - T_{abec} + S_{bcae} - S_{baec}\bigr) \notag \\
	&\phantom{= Q^{bc}\biggl(}+ \delta^a_bL_{ca} - \frac{1}{2}Q^{ae}(\partial_aQ_{bc})L_e - \frac{1}{2}Q_{bc}(\partial_aQ^{ae})L_e - \frac{1}{2}Q^{ae}Q_{bc}L_{ae}\biggr),
	\label{}
\end{align}
where we have used the symmetry of $Q^{bc}$ to interchange the indices $b$ and $c$ in several of the terms coming from \Eq{dG(E)}. Now a straightforward calculation, which relies on the identities
\begin{align}
	E^a_i\partial_bE_a^j &= -E_a^j\partial_bE^a_i \label{EE-1} \\
	E^a_i\partial_bE_c^i &= -E_c^i\partial_bE^a_i \label{EE-2}
\end{align}
as well as
\begin{equation}
	Q^{ab}Q_{bc} = \delta^a_c
	\label{}
\end{equation}
leads to the result
\begin{align}
	E^a_iE^b_i\partial_c\Gamma^c_{ab} &= 2Q^{ab}U{}\updown{c}{cab} + Q^{ab}Q^{cd}\bigl(T_{acdb} - T_{acbd}\bigr) \notag \\
	&\quad + 2Q^{ab}S\updown{c}{cab} - Q^{ab}S\updown{c}{abc} - (\partial_aQ^{ab})C\updown{c}{bc} - (\partial_aE^a_i)(\partial_bE^b_i) \notag \\
	&\quad - Q^{ab}C\updown{c}{ac}L_b - \frac{3}{2}(\partial_aQ^{ab})L_b - \frac{1}{2}Q^{ab}L_{ab}.
	\label{term1}
\end{align}
For the second term in \Eq{R3}, we need to evaluate
\begin{align}
	E^a_iE^b_i\partial_b\Gamma^c_{ac}& \notag \\
	= \frac{1}{2}Q^{ab}\biggl(&U{}\updown{c}{bac} + U{}\updown{c}{bca} + S\updown{c}{bac} + S\updown{c}{bca} + \partial_bQ^{ce}\bigl(C_{ace} + C_{cae} - C_{aec} - C_{cea}\bigr) \notag \\
&+ Q^{ce}\bigl(T_{bace} + T_{bcae} - T_{baec} - T_{bcea} + S_{acbe} + S_{cabe} - S_{abec} - S_{cbea}\bigr) \notag \\
&- Q^{ce}(\partial_bQ_{ac})L_e - Q_{ac}(\partial_bQ^{ce})L_e + 3L_{ab}\biggr).
	\label{}
\end{align}
Here several terms immediately cancel due to the symmetry of $Q^{ce}$. Then, after noting that
\begin{equation}
	\partial_bQ^{ce}\bigl(C_{cae} - C_{cea}\bigr) = U{}\updown{c}{bac} - U{}\updown{c}{bca} + Q^{ce}\bigl(T_{aebc} - T_{eabc}\bigr),
	\label{}
\end{equation}
we find that the remaining terms reduce to
\begin{equation}
	E^a_iE^b_i\partial_b\Gamma^c_{ac} = Q^{ab}U{}\updown{c}{abc} + Q^{ab}S\updown{c}{abc} + \frac{3}{2}Q^{ab}L_{ab}.
	\label{term2}
\end{equation}

\subsection{Ricci scalar. Terms with two Christoffel symbols}

We then move on to the third term in \Eq{R3}. First, a short calculation shows that
\begin{equation}
	E^a_iE^b_i\Gamma^c_{ab} = -Q^{ac}C\updown{b}{ab} - \partial_aQ^{ac} - \frac{1}{2}Q^{ac}L_a.
	\label{term3.1}
\end{equation}
Then contracting this with
\begin{equation}
	\Gamma^d_{cd} = C\updown{d}{cd} + \frac{3}{2}L_c,
	\label{term3.2}
\end{equation}
we obtain
\begin{equation}
	E^a_iE^b_i\Gamma^c_{ab}\Gamma^d_{cd} = -Q^{ab}C\updown{c}{ac}C\updown{d}{bd} - (\partial_aQ^{ab})C\updown{c}{bc} - 2Q^{ab}C\updown{c}{ac}L_b - \frac{3}{2}(\partial_aQ^{ab})L_b - \frac{3}{4}Q^{ab}L_aL_b.
	\label{term3}
\end{equation}
Now it remains to deal with the last term in \Eq{R3}. Let us again start by considering the contraction of the metric with one of the Christoffel symbols, namely $E^a_iE^b_i\Gamma^c_{ad}$. Since this will be contracted with $\Gamma^d_{bc}$, which is symmetric in $b$ and $c$, it is sufficient to keep only the symmetric part of the expression, which turns out to be
\begin{equation}
	E^a_iE^{(b}_i\Gamma^{c)}_{ad} = Q_{\phantom{ad}}^{a(b}C\updown{c)}{da} + \frac{1}{2}Q^{bc}L_d.
	\label{}
\end{equation}
Now contracting this with
\begin{equation}
	\Gamma^d_{bc} = \frac{1}{2}\Bigl(C\updown{d}{bc} + C\updown{d}{cb} + Q^{de}\bigl(C_{bce} + C_{cbe} - C_{bec} - C_{ceb}\bigr) + \delta^d_bL_c + \delta^d_cL_b  - Q^{de}Q_{bc}L_e\Bigr),
	\label{}
\end{equation}
carefully multiplying out all the terms and carrying out a number of additional simplifications, we eventually arrive at the result
\begin{align}
	E^a_iE^b_i\Gamma^c_{ad}\Gamma^d_{bc} &= \frac{1}{2}Q^{ab}\bigl(U{}\updown{c}{abc} - U{}\updown{c}{acb}\bigr) + \frac{1}{2}Q^{ab}Q^{cd}\bigl(T_{acdb} - T_{acbd}\bigr) \notag \\
	&\quad - \frac{1}{2}A\updown{ab}{c}C\updown{c}{ab} + \frac{1}{2}(\partial_aE^b_i)(\partial_bE^a_i) \notag \\
	&\quad - (\partial_aQ^{ab})L_b - Q^{ab}C\updown{c}{ac}L_b - \frac{1}{4}Q^{ab}L_aL_b.
	\label{term4}
\end{align}

\subsection{Expression in terms of the triad and its partial derivatives}

By combining \Eqs{term1}, \eqref{term2}, \eqref{term3} and \eqref{term4}, we obtain the following expression for the Ricci scalar as a function of the densitized triad and its partial derivatives:
\begin{align}
	&\de\Rt \; = \; 2Q^{ab}\bigl(S\updown{c}{cab} - S\updown{c}{abc}\bigr) \notag \\
&\quad + 2Q^{ab}U{}\updown{c}{cab} + \frac{1}{2}Q^{ab}U{}\updown{c}{acb} - \frac{3}{2}Q^{ab}U{}\updown{c}{abc} + \frac{1}{2}Q^{ab}Q^{cd}\bigl(T_{acdb} - T_{acbd}\bigr) \notag \\
&\quad + \frac{1}{2}A\updown{ab}{c}C\updown{c}{ab} - Q^{ab}C\updown{c}{ac}C\updown{d}{bd} - 2(\partial_aQ^{ab})C\updown{c}{bc} - (\partial_aE^a_i)(\partial_bE^b_i) - \frac{1}{2}(\partial_aE^b_i)(\partial_bE^a_i) \notag \\
&\quad - 2Q^{ab}C\updown{c}{ac}L_b - 2(\partial_aQ^{ab})L_b - \frac{1}{2}Q^{ab}L_aL_b - 2Q^{ab}L_{ab}.
	\label{result1}
\end{align}
Before continuing to the next stage of our calculation, we perform one more series of manipulations, which consists of using the identities \eqref{EE-1} and \eqref{EE-2} to transform derivatives of the inverse triad into derivatives of the triad itself whenever this is possible. This has the advantage of slightly reducing the number of different terms appearing in the final expression for the Ricci scalar; in particular, second derivatives of the inverse triad will be completely removed from the expression.

For example, the first term of \Eq{result1} can be rewritten as follows:
\begin{align}
	Q^{ab}S\updown{c}{cab} &= E^a_iE^b_iE^c_j\partial_c\partial_aE_b^j \notag \\
	&= -E^a_iE^b_i\Bigl(E_b^j\partial_c\partial_aE^c_j + (\partial_aE_b^j)(\partial_cE^c_j) + (\partial_aE^c_j)(\partial_cE_b^j)\Bigr) \notag \\
	&= -E^a_i\partial_a\partial_bE^b_i + E^a_i(\partial_aE^b_i)E_b^j(\partial_cE^c_j) + E^a_i(\partial_cE^b_i)E_b^j(\partial_aE^c_j) \notag \\
	&= -E^a_i\partial_a\partial_bE^b_i + A\updown{ab}{a}B\downup{bc}{c} + A\updown{ab}{c}B\downup{ba}{c}
	\label{}
\end{align}
After the entire expression has been treated in this way, and the derivatives of $Q^{ab}$ are written as $\partial_cQ^{ab} = A\updown{ab}{c} + A\updown{ba}{c}$, we are left with the result
\begin{align}
	\de\Rt = & -2E^a_i\partial_a\partial_bE^b_i + 2Q^{ab}E_c^i\partial_a\partial_bE^c_i \notag \\
	& - (\partial_a E^a_i)(\partial_bE^b_i) - \frac{1}{2}(\partial_aE^b_i)(\partial_bE^a_i) \notag \\
	& + \frac{5}{2}Q^{ab}(\partial_aE^c_i)(\partial_bE_c^i) - \frac{1}{2}Q^{ab}Q_{cd}(\partial_aE^c_i)(\partial_bE^d_i) \notag \\
	& + 2A\updown{ab}{a}B\downup{cb}{c} + 2A\updown{ab}{b}B\downup{ca}{c} + A\updown{ab}{c}B\downup{ba}{c} \notag \\
	& + \frac{1}{2}Q_{ab}A\updown{ca}{d}A\updown{db}{c} - Q^{ab}B\downup{ca}{c}B\downup{db}{d} \notag \\
	& + 2\bigl(Q^{ab}B\downup{ca}{c} - A\updown{ab}{a} - A\updown{ba}{a}\bigr)L_b - \frac{1}{2}Q^{ab}L_aL_b - 2Q^{ab}L_{ab}.
	\label{result2}
\end{align}

\subsection{Replacing partial derivatives with covariant derivatives}
\label{sec:replacing}

In \Eq{result2} we have obtained an expression which gives the Ricci scalar in terms of the densitized triad and its partial derivatives. However, our eventual goal was to write the Ricci scalar as a function of the triad and its gauge covariant derivatives. The gauge covariant derivative of the densitized triad is defined by
\begin{equation}
	\D_aE^b_i = \partial_aE^b_i + \epsilon\downup{ij}{k}A_a^jE^b_k.
	\label{DE_i}
\end{equation}
Since the covariant derivative transforms covariantly under internal gauge transformations, we can consistently apply the definition \eqref{DE_i} to the covariant derivative itself. This leads to the expression
\begin{equation}
	\D_a\D_bE^c_i = \partial_a\partial_bE^c_i + \epsilon\downup{ij}{k}\Bigl(\partial_a\bigl(A_b^jE^c_k\bigr) + A_a^j\bigl(\partial_bE^c_k\bigr)\Bigr) + \bigl(A_a^jE^c_j)A_{bi} - \bigl(A_a^jA_b^j\bigr)E^c_i.
	\label{DDE_i}
\end{equation}
for the second covariant derivative of the triad.

We now proceed to use \Eqs{DE_i} and \eqref{DDE_i} to express all partial derivatives of the triad in \Eq{result2} in terms of the corresponding gauge covariant derivatives. For first derivatives of the triad, we have
\begin{equation}
	\partial_aE^b_i = \D_aE^b_i - \epsilon\downup{ij}{k}A_a^jE^b_k.
	\label{dE_i}
\end{equation}
To deal with the second derivatives, we solve \Eq{DDE_i} for $\partial_a\partial_bE^c_i$ and apply \Eq{dE_i} to the first derivatives of the triad, finding
\begin{align}
	\partial_a\partial_bE^c_i= \D_a\D_bE^c_i &- \epsilon\downup{ij}{k}\Bigl(\bigl(\partial_aA_b^j\bigr)E^c_k + A_a^j\bigl(\D_bE^c_k\bigr) + A_b^j\bigl(\D_aE^c_k\bigr)\Bigr) \notag \\
	&+ \bigl(A_b^jE^c_j\bigr)A_{ai} - \bigl(A_a^jA_b^j\bigr)E^c_i.
	\label{}
\end{align}
Finally, we add this expression to itself with the indices $a$ and $b$ interchanged to obtain
\begin{align}
	2\partial_a\partial_bE^c_i &= \bigl(\D_a\D_b + \D_b\D_a\bigr)E^c_i - 2\epsilon\downup{ij}{k}\Bigl( A_a^j\bigl(\D_bE^c_k\bigr) + A_b^j\bigl(\D_aE^c_k\bigr)\Bigr) \notag \\
	&\qquad - \epsilon\downup{ij}{k}\bigl(\partial_aA_b^j + \partial_bA_a^j\bigr)E^c_k + \bigl(A_a^jE^c_j\bigr)A_{bi} + \bigl(A_b^jE^c_j\bigr)A_{ai} - 2\bigl(A_a^jA_b^j\bigr)E^c_i,
	\label{ddE_i}
\end{align}
where the symmetric part of the second covariant derivative appears on the right-hand side.

We now substitute \Eq{dE_i} for every first derivative of the triad in \Eq{result2} and \Eq{ddE_i} for every second derivative. (Note that derivatives of $\de$ do not need to be substituted with covariant derivatives, since the determinant is a gauge invariant object, and therefore its gauge covariant derivative coincides with the partial derivative.) After a long but comparatively straightforward calculation, we find that all the additional terms generated by the substitution cancel out among each other, provided that second derivatives are substituted with the expression \eqref{ddE_i} featuring the symmetric part of the second covariant derivative.

As an illustration, consider the term
\begin{equation}
	2\bigl(Q^{ab}B\downup{ca}{c} - A\updown{ab}{a} - A\updown{ba}{a}\bigr)L_b
	\label{}
\end{equation}
on the last line of \Eq{result2}. Since this is the only term containing exactly one factor of $L_b$, the correction terms arising from this term must cancel out among themselves. Performing the substitution indicated by \Eq{dE_i}, we indeed find
\begin{align}
	&Q^{ab}B\downup{ca}{c} - A\updown{ab}{a} - A\updown{ba}{a} \notag \\
	&= Q^{ab}\bigl(E_c^i\partial_aE^c_i\bigr) - E^a_i\partial_aE^b_i - E^b_i\partial_aE^a_i \notag \\
	&= Q^{ab}E_c^i\bigl(\D_aE^c_i - \epsilon\downup{ij}{k}A_a^jE^c_k\bigr) - E^a_i\bigl(\D_aE^b_i - \epsilon\downup{ij}{k}A_a^jE^b_k\bigr) - E^b_i\bigl(\D_aE^a_i - \epsilon\downup{ij}{k}A_a^jE^a_k\bigr) \notag \\
	&= Q^{ab}\bigl(E_c^i\D_aE^c_i\bigr) - E^a_i\D_aE^b_i - E^b_i\D_aE^a_i - Q^{ab}\epsilon\downup{jk}{l}A_a^k\delta^j_l + \epsilon\downup{ij}{k}A_a^j\bigl(E^a_iE^b_k + E^a_kE^b_i\bigr) \notag \\
	&= Q^{ab}\bigl(E_c^i\D_aE^c_i\bigr) - E^a_i\D_aE^b_i - E^b_i\D_aE^a_i.
	\label{}
\end{align}
Thus, our conclusion is that the Ricci scalar is given as a function of the densitized triad and its gauge covariant derivatives by the following expression:
\begin{align}
	\de\Rt = & -2E^a_i\D_{(a}\D_{b)}E^b_i + 2Q^{ab}E_c^i\D_a\D_bE^c_i \notag \\
	& - (\D_a E^a_i)(\D_bE^b_i) - \frac{1}{2}(\D_aE^b_i)(\D_bE^a_i) \notag \\
	& + \frac{5}{2}Q^{ab}(\D_aE^c_i)(\D_bE_c^i) - \frac{1}{2}Q^{ab}Q_{cd}(\D_aE^c_i)(\D_bE^d_i) \notag \\
	& + 2\A\updown{ab}{a}\B\downup{cb}{c} + 2\A\updown{ab}{b}\B\downup{ca}{c} + \A\updown{ab}{c}\B\downup{ba}{c} \notag \\
	& + \frac{1}{2}Q_{ab}\A\updown{ca}{d}\A\updown{db}{c} - Q^{ab}\B\downup{ca}{c}\B\downup{db}{d} \notag \\
	& + 2\bigl(Q^{ab}\B\downup{ca}{c} - \A\updown{ab}{a} - \A\updown{ba}{a}\bigr)C_b + \frac{3}{2}Q^{ab}C_aC_b - 2Q^{ab}C_{ab} \label{}
\end{align}
Here we have passed from the logarithmic derivative $L_{ab}$ to the variable
\begin{equation}
	C_{ab} = \frac{\partial_a\partial_b\de}{\de}
	\label{}
\end{equation}
introduced in \Eq{C_ab} (and $L_a$ has been renamed to $C_a$ for consistency), and we have introduced the new abbreviations
\begin{align}
	\A\updown{ab}{c} &= E^a_i\D_c E^b_i \\
	\B\downup{ab}{c} &= E_a^i\D_b E^c_i.
\end{align}

\section{Regularization of covariant derivatives of the triad}
\label{sec:B}

In this Appendix we present in detail the calculations which confirm the validity of the regularized variables introduced in section \ref{sec:derivatives} to represent covariant derivatives of the densitized triad. In sections \ref{sec:h_line} and \ref{sec:square} we begin by deriving some auxiliary results which will be used in the main calculations. In sections \ref{DaE proof}--\ref{DabE proof} we then establish that the expressions \eqref{DaE}, \eqref{DaaEb} and \eqref{DabEc} do provide correct regularizations respectively for the first covariant derivative $\D_aE^b$, the pure second derivative $\D_a^2E^b$ and the symmetric part of the mixed second derivative $\D_a\D_bE^c$. 

\subsection{Holonomy along an infinitesimal line segment}
\label{sec:h_line}

In preparation for the evaluation of parallel transported flux variables in the following section, we will compute the holonomy of the Ashtekar connection along a straight line segment of infinitesimal coordinate length $\epsilon$. For any path $e$, the holonomy $h_e$ is defined by the expression
\begin{align}
	h_e &= {\cal P}\exp\biggl(-\int_e A\biggr) \notag \\
	&= \sum_n (-1)^n \int_0^1 ds \int_0^{s_1} ds_2 \cdots \int_0^{s_{n-1}}ds_n\,\dot e^{a_1}(s_1)\cdots \dot e^{a_n}(s_n)A_{a_1}\bigl(e(s_1)\bigr)\cdots A_{a_n}\bigl(e(s_n)\bigr),
	\label{holonomy}
\end{align}
where $A_a = A_a^i\tau_i$, and the path $e^a(s)$ is parametrized by a parameter $s$ running from $0$ to $1$. In our case, the natural parametrization of the line segment is given by
\begin{align}
	e^a(s) &= s\epsilon u^a \\
	\dot e^a(s) &= \epsilon u^a
	\label{}
\end{align}
with $u^a$ being the (constant) unit tangent vector of the line segment.

For our purposes, it suffices to evaluate \Eq{holonomy} up to terms of order $\epsilon^2$. Truncating the series at the second order, we have
\begin{align}
	h_e = \Id - \int_0^1 ds\,\dot e^a(s) A_a\bigl(e(s)\bigr) + \int_0^1 ds\int_0^s dt\,\dot e^a(s)\dot e^b(t)A_a\bigl(e(s)\bigr)A_b\bigl(e(t)\bigr) + {\cal O}\bigr(\epsilon^3\bigl).
	\label{h_2nd_order}
\end{align}
In the first integral, the connection can be expanded around the beginning point of the line segment as
\begin{equation}
	A_a\bigl(e(s)\bigr) = A_a(0) + e^b(s)\partial_bA_a(0) + {\cal O}\bigl(\epsilon^2\bigr) = A_a(0) + \epsilon s u^b\partial_bA_a(0) + {\cal O}\bigl(\epsilon^2\bigr).
	\label{}
\end{equation}
In this way we find
\begin{align}
	\int_0^1 ds\,\dot e^a(s)A_a\bigl(e(s)\bigr) = \epsilon u^a A_a(0) + \frac{1}{2}\epsilon^2u^au^b\partial_bA_a(0) + {\cal O}\bigl(\epsilon^3\bigr).
	\label{}
\end{align}
In the double integral in \Eq{h_2nd_order}, at order $\epsilon^2$ the connection can simply be replaced with its value at $0$, giving
\begin{align}
	&\int_0^1 ds\int_0^s dt\,\dot e^a(s)\dot e^b(t)A_a\bigl(e(s)\bigr)A_b\bigl(e(t)\bigr) = \frac{1}{2}\epsilon^2u^au^bA_a(0)A_b(0) + {\cal O}\bigl(\epsilon^3\bigr).
	\label{}
\end{align}
Hence we have shown that
\begin{equation}
	h_e = \Id - \epsilon u^aA_a(0) + \frac{1}{2}\epsilon^2 u^au^b\Bigl(A_a(0)A_b(0) - \partial_bA_a(0)\Bigr) + {\cal O}\bigl(\epsilon^3\bigr).
	\label{h_l}
\end{equation}
This result can be expressed in a slightly more compact form as follows: Let
\begin{equation}
	\xi^a = \frac{1}{2}\epsilon u^a
	\label{}
\end{equation}
denote the midpoint of the line segment. Then we have
\begin{equation}
	A_a(\xi) = A_a(0) + \xi^b \partial_bA_a(0) + {\cal O}\bigl(\epsilon^2\bigr)= A_a(0) + \frac{1}{2}\epsilon u^b\partial_bA_a(0) + {\cal O}\bigl(\epsilon^2\bigr)
	\label{}
\end{equation}
and comparing with \Eq{h_l}, we see that we can write
\begin{equation}
	h_e = \Id - \epsilon u^aA_a(\xi) + \frac{1}{2}\epsilon^2u^au^bA_a(\xi)A_b(\xi) + {\cal O}\bigl(\epsilon^3\bigr).
	\label{h_line}
\end{equation}
Note also that the inverse holonomy is given by
\begin{equation}
	h_e^{-1} = \Id + \epsilon u^aA_a(\xi) + \frac{1}{2}\epsilon^2u^au^bA_a(\xi)A_b(\xi) + {\cal O}\bigl(\epsilon^3\bigr),
	\label{h_inv}
\end{equation}
since the inverted path $e^{-1}$ has the opposite tangent vector $-u^a$ and the same midpoint $\xi^a$ as the original path.

\subsection{Parallel transported flux through an infinitesimal square surface}
\label{sec:square}

\begin{figure}[t]
	\centering
	\includegraphics[scale=0.17]{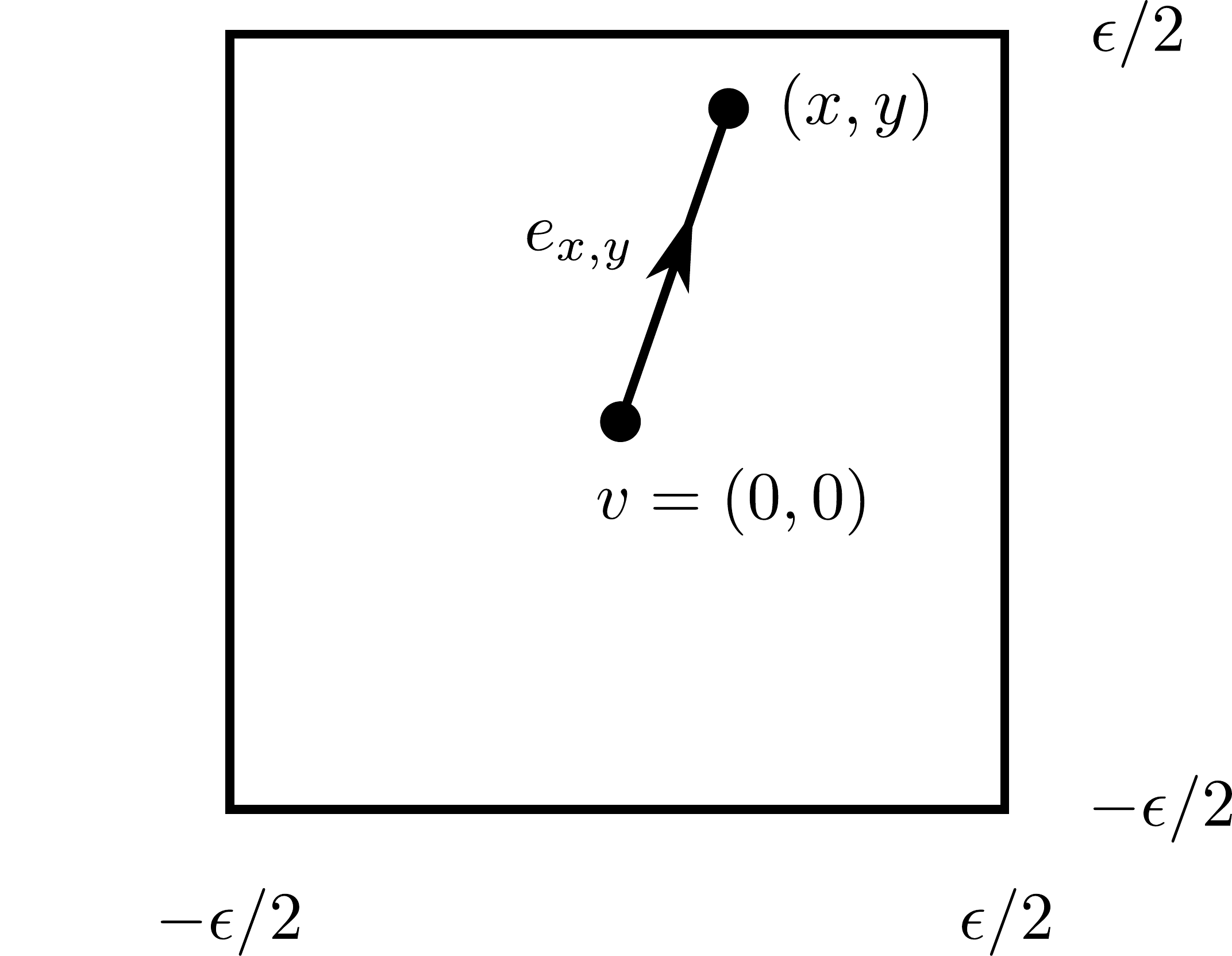}
	\caption{Parallel transported flux variable associated to an infinitesimal square surface. Straight-line paths are used to perform parallel transport between the midpoint $v$ and a given point $(x, y)$ on the surface.}
	\label{fig:square}
\end{figure}

The parallel transported flux variable, introduced in section \ref{sec:transported}, is defined by the expressions
\begin{equation}
	\Et_i(S, x_0) = -2\Tr\Bigl(\tau_i\Et(S, x_0)\Bigr)
	\label{}
\end{equation}
\begin{equation}
	\Et(S, x_0) = \int_S d^2\sigma\,n_a(\sigma)h_{x_0,x(\sigma)}E^a\bigl(x(\sigma)\bigr)h_{x_0,x(\sigma)}^{-1}
	\label{}
\end{equation}
where $E^a = E^a_i\tau^i$ is the densitized triad, and $h_{x_0, x(\sigma)}$ are holonomies associated to a family of paths $p_{x(\sigma)\to x_0}$ which connect each point on the surface $S$ to a fixed point $x_0$ (which may lie on the surface or outside of it).

We will evaluate the parallel transported flux associated to a square-shaped surface of infinitesimal coordinate area $\epsilon^2$, with the parallel transport taken to the midpoint of the surface along straight lines connecting each point on the surface to the midpoint. Choosing the surface -- which we denote by $S^z$ -- to lie in the $xy$-coordinate plane of a coordinate system whose origin coincides with the center of the surface, we have to compute the integral
\begin{equation}
	\Et\bigl(S^z(v), v\bigr) = \int_{-\epsilon/2}^{\epsilon/2} dx \int_{-\epsilon/2}^{\epsilon/2} dy\,h_{x, y}^{-1}E^z(x, y)h_{x, y}
	\label{square-int}
\end{equation}
where $h_{x, y} \equiv h_{e_{x, y}}$, with $e_{x, y}$ being a straight-line path from the midpoint $v = (0, 0)$ to the point $(x, y)$ as shown in \Fig{fig:square}.

Our goal is to evaluate the integral \eqref{square-int} up to terms of order $\epsilon^4$. Since each integral contributes one power of $\epsilon$, the integrand must be expanded to the second order. By \Eq{h_line}, the holonomy $h_{x, y}$ is
\begin{equation}
	h_{x, y} = \Id - \delta u^aA_a(\xi) + \frac{1}{2}\delta^2u^au^bA_a(\xi)A_b(\xi) + {\cal O}\bigl(\delta^3\bigr)
	\label{h_xy}
\end{equation}
with
\begin{equation}
	\delta = \sqrt{x^2 + y^2}
	\label{}
\end{equation}
the coordinate length of the path $e_{x, y}$,
\begin{equation}
	u^a = \frac{1}{\sqrt{x^2 + y^2}}(x, y)
	\label{}
\end{equation}
the unit tangent vector, and
\begin{equation}
	\xi^a = \bigl(x/2, y/2\bigr)
	\label{}
\end{equation}
the midpoint of the path. Thus,
\begin{align}
	h_{x, y} = \Id &- xA_x(\xi) - yA_y(\xi) + \frac{1}{2}\Bigl(x^2A_x^2 + y^2A_y^2 + xy\bigl(A_xA_y + A_yA_x\bigr)\Bigr)\bigg|_{\xi} + {\cal O}\bigl(\delta^3\bigr)
	\label{}
\end{align}
and expanding the connection around the point $v = (0, 0)$ as
\begin{equation}
	A_a(\xi) = A_a(x/2, y/2) = A_a(v) + \frac{x}{2}\partial_xA_a(v) + \frac{y}{2}\partial_yA_a(v) + {\cal O}\bigl(\delta^2\bigr)
	\label{}
\end{equation}
we obtain
\begin{align}
	h_{x, y} = \Id &- xA_x(v) - yA_y(v) \notag \\
	&+ \frac{x^2}{2}\Bigl(A_x^2(v) - \partial_xA_x(v)\Bigr) + \frac{y^2}{2}\Bigl(A_y^2(v) - \partial_yA_y(v)\Bigr) \notag \\
	&+ \frac{xy}{2}\Bigl(A_x(v)A_y(v) + A_y(v)A_x(v) - \partial_xA_y(v) - \partial_yA_x(v)\Bigr) + {\cal O}\bigl(\delta^3\bigr).
	\label{}
\end{align}
In the same way, using \Eq{h_inv}, the inverse holonomy $h_{x, y}^{-1}$ is
\begin{align}
	h_{x, y}^{-1} = \Id &+ xA_x(v) + yA_y(v) \notag \\
	&+ \frac{x^2}{2}\Bigl(A_x^2(v) + \partial_xA_x(v)\Bigr) + \frac{y^2}{2}\Bigl(A_y^2(v) + \partial_yA_y(v)\Bigr) \notag \\
	&+ \frac{xy}{2}\Bigl(A_x(v)A_y(v) + A_y(v)A_x(v) + \partial_xA_y(v) + \partial_yA_x(v)\Bigr) + {\cal O}\bigl(\delta^3\bigr)
	\label{}
\end{align}
while the expansion of the densitized triad to second order reads
\begin{align}
	E^z(x, y) = E^z(v) &+ x\partial_xE^z(v) + y\partial_yE^z(v) \notag \\
	&+ \frac{x^2}{2}\partial_x^2E^z(v) + \frac{y^2}{2}\partial_y^2E^z(v) + xy\partial_x\partial_yE^z(v) + {\cal O}\bigl(\delta^3\bigr).
	\label{}
\end{align}
Inserting these expressions into \Eq{square-int} and collecting the terms up to order $\epsilon^4$, we find
\begin{align}
	\Et\bigl(S^z(v), v\bigr) = \epsilon^2E^z(v) + \frac{\epsilon^4}{24}\biggl(&\partial_x^2E^z + 2\bigl[A_x, \partial_xE^z\bigr] + \bigl[\partial_xA_x, E^z\bigr] \notag \\
	&+ \partial_y^2E^z + 2\bigl[A_y, \partial_yE^z\bigr] + \bigl[\partial_yA_y, E^z \bigr] \phantom{\Big|} \notag \\
	&+ A_x^2E^z - 2A_xE^zA_x + E^zA_x^2 \phantom{\Big|} \notag \\
	&+ A_y^2E^z - 2A_yE^zA_y + E^zA_y^2\biggr)\bigg|_v + {\cal O}\bigl(\epsilon^5\bigr).
	\label{}
\end{align}
Now recognizing that
\begin{align}
	&A_a^2E^z - 2A_aE^zA_a + E^zA_a^2 = \Bigl[A_a, \bigl[A_a, E^z\bigr]\Bigr] 
	\label{}
\end{align}
and comparing with the definition of the second covariant derivative given by \Eq{DDE}, we see that our result can be expressed as
\begin{equation}
	\Et\bigl(S^z(v), v\bigr) = \epsilon^2E^z(v) + \frac{\epsilon^4}{24}\Bigl(\D_x^2E^z(v) + \D_y^2E^z(v)\Bigr) + {\cal O}\bigl(\epsilon^5\bigr).
	\label{E-square}
\end{equation}

\subsection{First derivatives}
\label{DaE proof}

In section \ref{sec:derivatives}, the variable
\begin{equation}
	\Delta_aE\bigl(S^b, v\bigr) = \frac{\Et\bigl(S^b(v_a^+), v\bigr) - \Et\bigl(S^b(v_a^-), v\bigr)}{2}
	\label{DaE_}
\end{equation}
was introduced to regularize the covariant derivative $\D_aE^b(v)$. The surfaces and nodes involved in the regularization are illustrated by \Fig{fig:dE}, with $v_a^+$ and $v_a^-$ denoting the nodes immediately following and preceding the central node $v$ in the direction of the $x^a$-coordinate axis.

We will now expand the right-hand side of \Eq{DaE_} in powers of the regularization parameter $\epsilon$ in order to verify that the leading term of the expansion does reproduce the covariant derivative $\D_aE^b$ evaluated at $v$. Letting $e_a^+$ and $e_a^-$ denote the edges which connect the nodes $v_a^+$ and $v_a^-$ to the central node $v$ (the orientation of the edges agreeing with the positive direction of the $x^a$-coordinate axis as shown in \Fig{fig:dE}), the parallel transported flux variables entering \Eq{DaE} can be written as
\begin{align}
	\Et\bigl(S^b(v_a^+), v\bigr) &= h_{e_a^+}^{-1}\Et\bigl(S^b(v_a^+), v_a^+\bigr)h_{e_a^+} \label{ES+_} \\[1ex]
	\Et\bigl(S^b(v_a^-), v\bigr) &= h_{e_a^-}\Et\bigl(S^b(v_a^-), v_a^-\bigr)h_{e_a^-}^{-1} \label{ES-_}
\end{align}
where the flux variables on the right-hand side are of the form considered in section \ref{sec:square}. Applying now \Eqs{h_line} and \eqref{h_inv} to the edges $e_a^\pm$, and truncating the expansion at linear order in $\epsilon$, we find
\begin{align}
	h_{e_a^+} &= \Id - \epsilon A_a(v) + {\cal O}\bigl(\epsilon^2\bigr) \label{b1} \\
	h_{e_a^+}^{-1} &= \Id + \epsilon A_a(v) + {\cal O}\bigl(\epsilon^2\bigr)
\end{align}
and
\begin{align}
	h_{e_a^-} &= \Id - \epsilon A_a(v) + {\cal O}\bigl(\epsilon^2\bigr) \\
	h_{e_a^-}^{-1} &= \Id + \epsilon A_a(v) + {\cal O}\bigl(\epsilon^2\bigr).
\end{align}
Moreover, using \Eq{E-square},
\begin{align}
	\Et\bigl(S^b(v_a^\pm), v_a^\pm\bigr) = \epsilon^2E^b(v_a^\pm) + {\cal O}\bigl(\epsilon^4\bigr) = \epsilon^2E^b(v) \pm \epsilon^3\partial_aE^b(v) + {\cal O}\bigl(\epsilon^4\bigr).
	\label{e1}
\end{align}
Inserting then \Eqs{b1}--\eqref{e1} into \Eqs{ES+_} and \eqref{ES-_}, we see that
\begin{equation}
	\Et\bigl(S^b(v^a_\pm), v\bigr) = \epsilon^2E^b(v) \pm \epsilon^3\Bigl(\partial_aE^b(v) + \bigl[A_a(v), E^b(v)\bigr]\Bigr) + {\cal O}\bigl(\epsilon^4\bigr).
	\label{}
\end{equation}
Inside the parentheses we now have the covariant derivative $\D_aE^b(v)$. Hence we arrive at the conclusion
\begin{equation}
	\Delta_aE\bigl(S^b(v)\bigr) = \epsilon^3\D_aE^b(v) + {\cal O}\bigl(\epsilon^4\bigr),
	\label{}
\end{equation}
which confirms the expression \eqref{DaE_} as a valid regularization of the covariant derivative $\D_aE^b$ at $v$.

\subsection{Pure second derivatives}
\label{DaaE proof}

To regularize the pure second derivative $\D_a^2E^b$ at $v$, we introduced in \Eq{DaaEb} the variable
\begin{equation}
	\Delta_{aa}E\bigl(S^b, v\bigr) = \Et\bigl(S^b(v_a^+), v\bigr) - 2\Et\bigl(S^b(v), v\bigr) + \Et\bigl(S^b(v_a^-), v\bigr),
	\label{DaaEb_}
\end{equation}
which uses the same basic setup as the regularization of first derivatives, but now also involves the central node $v$. To verify the validity of the proposed regularization, we proceed as in the previous section; however, now the second-order correction terms must also be taken into account when expanding the variables involved in the right-hand side of \Eq{DaaEb_}. From \Eq{E-square}, we have
\begin{equation}
	\Et\bigl(S^b(v), v\bigr) = \epsilon^2E^b(v) + \frac{\epsilon^4}{24}\D_{\!\perp}^2(v) + {\cal O}\bigl(\epsilon^5\bigr),
	\label{b2}
\end{equation}
where we have introduced the abbreviation
\begin{equation}
	\D_{\!\perp}^2(v) = \sum_{a\neq b} \D_a^2E^b(v),
	\label{}
\end{equation}
and
\begin{equation}
	\Et\bigl(S^b(v_a^\pm), v_a^\pm\bigr) = \epsilon^2E^b(v) \pm \epsilon^3\partial_aE^b(v) + \frac{\epsilon^4}{2}\partial_a^2E^b(v) + \frac{\epsilon^4}{24}\D_{\!\perp}^2(v) + {\cal O}\bigl(\epsilon^5\bigr).
	\label{}
\end{equation}
Moreover, \Eqs{h_line} and \eqref{h_inv} show that the holonomies connecting $v_a^+$ and $v_a^-$ to $v$ are given by
\begin{align}
	h_{e_a^+} &= \Id - \epsilon A_a(v) + \frac{1}{2}\epsilon^2\Bigl(A_a^2(v) - \partial_aA_a(v)\Bigr) +  {\cal O}\bigl(\epsilon^3\bigr) \\
	h_{e_a^+}^{-1} &= \Id + \epsilon A_a(v) + \frac{1}{2}\epsilon^2\Bigl(A_a^2(v) + \partial_aA_a(v) \Bigr) + {\cal O}\bigl(\epsilon^3\bigr)
	\label{}
\end{align}
and
\begin{align}
	h_{e_a^-} &= \Id - \epsilon A_a(v) + \frac{1}{2}\epsilon^2\Bigl(A_a^2(v) + \partial_aA_a(v)\Bigr) + {\cal O}\bigl(\epsilon^3\bigr) \\
	h_{e_a^-}^{-1} &= \Id + \epsilon A_a(v) + \frac{1}{2}\epsilon^2\Bigl(A_a^2(v) - \partial_aA_a(v) \Bigr) + {\cal O}\bigl(\epsilon^3\bigr).
	\label{e2}
\end{align}
When \Eqs{b2}--\eqref{e2} are now inserted into \Eq{DaaEb}, we obtain
\begin{align}
	\Delta_{aa}E\bigl(S^b, v\bigr) = \epsilon^4&\biggl(\partial_a^2E^b(v) + 2\bigl[A_a(v), E^b(v)\bigr] + \bigl[\partial_aA_a(v), E^b(v)\bigr] \notag \\
	&+ A_a^2(v)E^b(v) - 2A_a(v)E^b(v)A_a(v) + E^b(v)A_a^2(v)\biggr) + {\cal O}\bigl(\epsilon^5\bigr).
	\label{}
\end{align}
Comparing this with \Eq{DDE} defining the second covariant derivative $\D_a\D_bE^c$, and recognizing that the last three terms within the parentheses are equal to the double commutator $\bigl[A_a, [A_a, E^b]\bigr]$ at $v$, we see that we have arrived at the desired result:
\begin{equation}
	\Delta_{aa}E\bigl(S^b(v)\bigr) = \epsilon^4\D_a^2E^b(v) + {\cal O}\bigl(\epsilon^5\bigr).
	\label{}
\end{equation}

\subsection{Mixed second derivatives}
\label{DabE proof}

\begin{figure}[t]
	\centering
	\begin{subfigure}{0.45\textwidth}
		\centering
		\includegraphics[scale=0.17]{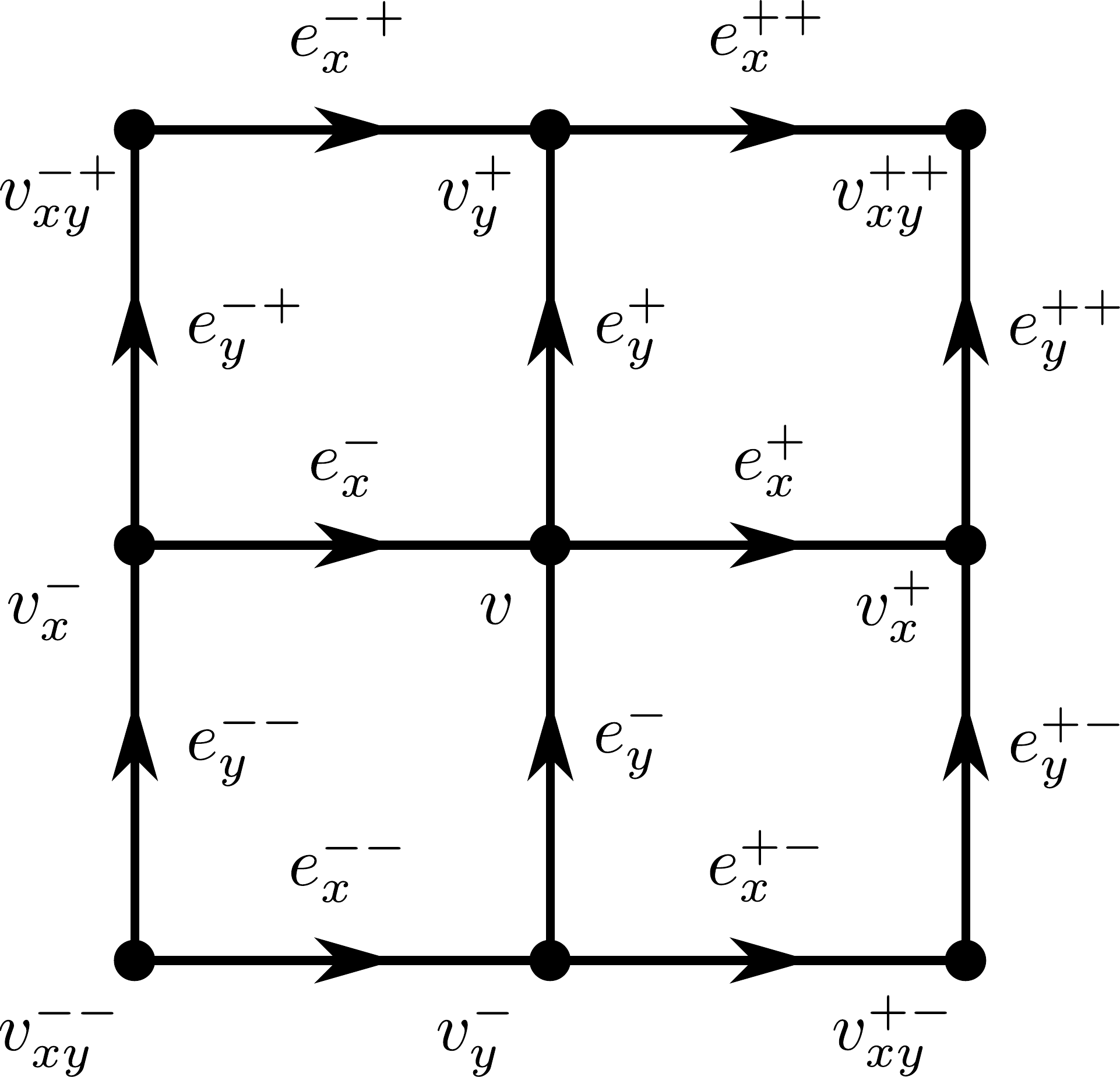}
	\end{subfigure}
	\begin{subfigure}{0.45\textwidth}
		\centering
		\begin{tabular}{CCC}
			e & u^a & \eta^a \\
			\hline
			e_x^+ & (1, 0) & (1/2, 0) \\
			e_x^- & (1, 0) & (-1/2, 0) \\
			e_y^+ & (0, 1) & (0, 1/2) \\
			e_y^- & (0, 1) & (0, -1/2) \\
			e_x^{++} & (1, 0) & (1/2, 1) \\
			e_x^{+-} & (1, 0) & (1/2, -1) \\
			e_x^{-+} & (1, 0) & (-1/2, 1) \\
			e_x^{--} & (1, 0) & (-1/2, -1) \\
			e_y^{++} & (0, 1) & (1, 1/2) \\
			e_y^{+-} & (0, 1) & (1, -1/2) \\
			e_y^{-+} & (0, 1) & (-1, 1/2) \\
			e_y^{--} & (0, 1) & (-1, -1/2) \\
		\end{tabular}
	\end{subfigure}
	\caption{The edges and nodes used to construct a regularization of the symmetric part of the second covariant derivative $\D_x\D_yE^z$ at $v$. The table on the right shows the unit tangent vector $u^a$ and the midpoint $\xi^a = \epsilon\eta^a$ for each edge involved in the construction.}
	\label{fig:labels}
\end{figure}

The regularization of the mixed second derivative $\D_a\D_bE^c(v)$ at the node $v$ uses the four nodes diagonally neighboring $v$ in the plane which contains $v$ and is spanned by the $x^a$- and $x^b$-coordinate directions of the background coordinate system (see \Fig{fig:labels}). We denote these nodes by $v_{ab}^{++}$, $v_{ab}^{+-}$, $v_{ab}^{-+}$ and $v_{ab}^{--}$. Furthermore, we introduce the symbol
\begin{equation}
	\sigma = ++, \; +-, \; {-+} \; \text{or} \; {--}
	\label{}
\end{equation}
labeling the four nodes, as well as the corresponding formal vector
\begin{equation}
	\sigma^a = \bigl(\sigma^1, \sigma^2\bigr)
	\label{}
\end{equation}
whose components are equal to $+1$ or $-1$ according to the value of the label $\sigma$; for example, if $\sigma = ++$, then $\sigma^a = (1, 1)$. With this notation, we define
\begin{equation}
	\Et\bigl(S^c(v_{ab}^{\sigma}), v\bigr)_{\rm sym.} = \frac{1}{2}\Bigl(\Et\bigl(S^c(v_{ab}^{\sigma}), v\bigr)_{v_{ab}^{\sigma} \to v_a^{\sigma^1} \to v} + \Et\bigl(S^c(v_{ab}^{\sigma}), v\bigr)_{v_{ab}^{\sigma} \to v_b^{\sigma^2} \to v}\Bigr)
	\label{}
\end{equation}
as a flux variable parallel transported symmetrically along the two available paths from $v_{ab}^\sigma$ to the central node $v$ (the subscripts on the right-hand side indicating the path used for the parallel transport in each of the flux variables). We then take
\begin{align}
	\Delta_{ab}E\bigl(S^c, v\bigr) \equiv \frac{1}{4}\biggl(\Et\bigl(&S^c(v_{ab}^{++}), v\bigr)_{\rm sym.} - \Et\bigl(S^c(v_{ab}^{+-}), v\bigr)_{\rm sym.} \notag \\
	&- \Et\bigl(S^c(v_{ab}^{-+}), v\bigr)_{\rm sym.} + \Et\bigl(S^c(v_{ab}^{--}), v\bigr)_{\rm sym.}\biggr)
	\label{DabEc_}
\end{align}
as the variable intended to regularize the symmetric part of the second covariant derivative $\D_a\D_bE^c$ at $v$.

To confirm that the variable \eqref{DabEc_} correctly approximates the symmetric part $\D_{(a}\D_{b)}E^c(v)$, we again use \Eqs{h_line}, \eqref{h_inv} and \eqref{E-square} to expand the right-hand side of \Eq{DabEc_} in powers of $\epsilon$. For each parallel transported flux variable in \Eq{DabEc_}, we can write
\begin{equation}
	\Et\bigl(S^z(v_{xy}^\sigma), v\bigr)_{v_{xy}^\sigma \to v_i^\sigma \to v} = \bigl(h_i^\sigma\bigr)^{-1}\Et\bigl(S^z(v_{xy}^\sigma), v_{xy}^\sigma\bigr)h_i^\sigma
	\label{E>>}
\end{equation}
where the flux variable on the right-hand side is of the form \eqref{E-square}, the subscript $i$ equals $a$ or $b$, and using the labels specified by \Fig{fig:labels}, the holonomies corresponding to the various possible values of $\sigma$ and $i$ are given by
\begin{align}
	h_a^{++} &= h_{e_b^{++}}h_{e_a^+} \label{b3} \\
	h_b^{++} &= h_{e_a^{++}}h_{e_b^+} \\
	h_a^{+-} &= h_{e_b^{+-}}^{-1}h_{e_a^+} \\
	h_b^{+-} &= h_{e_a^{+-}}h_{e_b^-}^{-1} \\
	h_a^{-+} &= h_{e_b^{-+}}h_{e_a^-}^{-1} \\
	h_b^{-+} &= h_{e_a^{-+}}^{-1}h_{e_b^+} \\
	h_a^{--} &= h_{e_b^{--}}^{-1}h_{e_a^-}^{-1} \\
	h_b^{--} &= h_{e_a^{--}}^{-1}h_{e_b^-}^{-1}. \label{e3}
\end{align}
For each edge shown in \Fig{fig:labels}, we use \Eqs{h_line} and \eqref{h_inv} to find the holonomy and inverse holonomy associated to the edge. Writing the midpoint of the edge as
\begin{equation}
	\xi^a = \epsilon\eta^a,
	\label{}
\end{equation}
we have
\begin{align}
	h_e &= \Id - \epsilon u^aA_a(v) + \frac{1}{2}\epsilon^2\Bigl(u^au^bA_a(v)A_b(v) - \eta^au^b\partial_aA_b(v)\Bigr) + {\cal O}\bigl(\epsilon^3\bigr), \label{h_2nd} \\[2ex]
	h_e^{-1} &= \Id + \epsilon u^aA_a(v) + \frac{1}{2}\epsilon^2\Bigl(u^au^bA_a(v)A_b(v) + \eta^au^b\partial_aA_b(v)\Bigr) + {\cal O}\bigl(\epsilon^3\bigr). \label{h^2nd}
\end{align}
With the help of \Eqs{h_2nd} and \eqref{h^2nd}, each of the holonomies \eqref{b3}--\eqref{e3} can now be expanded up to terms of order $\epsilon^2$. Moreover, applying \Eq{E-square} to the flux variable on the right-hand side of \Eq{E>>}, we obtain
\begin{align}
	\Et\bigl(S^c(v^\sigma), v^\sigma\bigr) &= \epsilon^2E^c(v^{\sigma}) + \frac{\epsilon^4}{24}\D_{\!\perp}^2(v^\sigma) + {\cal O}\bigl(\epsilon^5\bigr) \notag \\
	&= \epsilon^2E^c(v) + \epsilon^3\sigma^a\partial_aE^c(v) + \frac{1}{2}\epsilon^4\sigma^a\sigma^b\partial_a\partial_bE^c(v) + \frac{\epsilon^4}{24}\D_{\!\perp}^2(v) + {\cal O}\bigl(\epsilon^5\bigr).
	\label{}
\end{align}
We then insert all this into \Eq{DabEc_}, and eventually find
\begin{equation}
	\Delta_{ab}E\bigl(S^c(v)\bigr) = \frac{\epsilon^4}{2}\Bigl(\D_a\D_bE^c(v) + \D_b\D_aE^c(v)\Bigr) + {\cal O}\bigl(\epsilon^5\bigr),
	\label{}
\end{equation}
which confirms that we have indeed managed to construct a regularization which correctly approximates the symmetric part of the mixed second derivative $\D_a\D_bE^c$ at $v$.

\end{document}